# Tantalum as a base material for superconducting integrated circuits

Evgeniy V. Zikiy,[1,2] Nikita S. Smirnov,[1,2] Stanislav A. Kotenkov,[1] Ilya A. Stepanov,[1] Julia A. Agafonova,[1] Daria A. Moskaleva,[1] Aleksei R. Matanin,[1] Anton I. Ivanov,[1] Dmitry O. Moskalev,[1] Julia A. Kurochkina,[1] Aleksander V. Andriyash,[2] and Ilya A. Rodionov[1,2,*]

[1]Shukhov Labs, Quantum Park, Bauman Moscow State Technical University, Moscow, 105005, Russia
[2]Dukhov Automatics Research Institute (VNIIA), Moscow 127055, Russia
*email: irodionov@bmstu.ru

The performance of superconducting integrated circuits for quantum applications is fundamentally limited by material-related losses. Tantalum, as an emerging material for next-generation quantum circuits, has attracted considerable attention in recent years after demonstrating breakthrough performance in both superconducting microwave resonators and qubits. Concurrently, a growing body of work is devoted to the operation of tantalum-based circuits and related fabrication techniques. This interest is further stimulated by tantalum thin films polymorphism resulting in a variety of its crystalline structure, superconducting properties, coherence, etc. Furthermore, tantalum circuits exhibit distinctive features in cryogenic experiments, which have not been observed in aluminum- or niobium-based ones. In this review, we summarize the recent research of tantalum thin films growth and phase selection mechanisms on various substrates, key aspects of fabrication and performance of superconducting circuit, including a material first-principles theoretical study. In conclusion, we address a number of open issues, including the role of β-phase impurities, the effect of hydrofluoric acid solutions on chain characteristics, and the anomalous behavior of α-tantalum chains at cryogenic temperatures.

# Table of contents

# I. Introduction

Superconducting quantum processors[1–5] are typically architected from an ensemble of superconducting qubits[6,7], readout resonators, couplers, and control circuitry, which may be co-fabricated on a monolithic substrate or distributed across multiple dies in a flip-chip configuration[8–10]. Peripheral components, including parametric amplifiers[11–15] and quantum memory elements[16–19], are frequently implemented on ancillary chips. The fundamental functional layer of all such integrated circuits is a base metallization layer, commonly delineated via lithographic patterning and etching of a thin superconducting film of thickness ranging from 100 to 300 nm. This layer defines the primary circuit elements, including coplanar waveguide resonators[20,21], transmission lines, parallel-plate capacitors, ground planes, and associated features with critical dimensions spanning from several micrometers to hundreds of micrometers. While the full circuit stack may incorporate Josephson junctions (JJs)[22,23], air-bridge crossovers[24–26], interlayer dielectrics[27], and metallic bumps for flip-chip integration, the base layer constitutes the foundational plane of the superconducting circuit architecture.

The identification of an optimal material for the base layer remains an open challenge. In terms of superconducting parameters Nb outperforms other materials. However, the performance of Nb-based circuits is similar to that of aluminum-based circuits, even though the Al critical temperature ($T_C$) is several times lower than for Nb. Historically, Al and Nb have been the materials of choice for base layer deposition, though sporadic efforts have explored other superconductors such as Ta[28], Hf[29], MoRe[30], and Re[31] with limited success. The application of superconducting nitrides – including TiN[32–34], NbN[35–37], and NbTiN[38–40] – has yielded more favorable outcomes; however, the substantial kinetic inductance intrinsic to these compounds imposes constraints on their utility in qubit circuitry. In 2021, transmon qubits fabricated using a Ta base layer demonstrated a record $T_1$ energy relaxation time exceeding 300 μs[41]. Subsequent advances extended this benchmark to over 500 μs by 2022[42], culminating in the recent surpassing of the 1 ms threshold in 2025[43]. Concurrently, Ta-based superconducting resonators have attained unprecedented performance metrics, with low-power internal quality factors ($Qi_{LP}$) in coplanar resonators surpassing 10 million for the first time[43–45]. Thus, tantalum has emerged as the prevailing frontrunner in the current state of the art.

The pronounced performance characteristics of tantalum have precipitated substantial interest within the global research community. The preceding quinquennium has witnessed a proliferation of literature detailing both the performance characteristics of superconducting devices incorporating Ta base layers and the associated technology and characterization of Ta thin films. This Review endeavors to consolidate these developments.

## A. Article organization

In Section II, we discuss the physical and morphological properties of tantalum thin films. We present approaches to obtaining the desired tantalum phase and current hypotheses regarding the phase selection mechanism.

In Section III, we discuss fabrication methods. We have attempted to outline the full range of manufacturing methods available for Ta circuits. Specifically, we describe post-treatment options that are particularly relevant for the Ta base layer.

In Section IV, we discuss the claimed performance of tantalum-based quantum circuits, such as resonators and qubits. We analyze the underlying mechanisms that contribute to their exceptional coherent properties.

In Section V, we examine the current limitations and unresolved fabrication and operational challenges inherent to Ta-based circuits. We focus on the impact of β-phase impurities and the effect of hydrofluoric acid solutions on circuit performance, as well as the anomalous behavior of α-Ta resonators.

In conclusion, we summarize the sources for the high performance of Ta circuits, current manufacturing issues, and challenges in the characterization of Ta circuits.

## II. Tantalum thin films

Bulk tantalum predominantly crystallizes in the α-phase, adopting a body-centered cubic (BCC) lattice with parameter a = 3.303 Å[46]. The metastable β-phase was first realized in thin-film form in 1965 via both physical vapor deposition (PVD) and chemical vapor deposition (CVD) techniques[46]. These β-Ta films exhibits a tetragonal crystal structure with lattice parameters a = 10.211 Å and c = 5.306 Å[47]. The synthesis of bulk β-Ta remains challenging and has been achieved primarily through high-temperature electrochemical routes, such as electrolysis in fluoride melts at temperatures exceeding 1000 K[47,48]. In contrast, in the thin film form, Ta grows in the β-phase without additional treatments on such common substrates as silicon[49,50], quartz and glass[49,51], sapphire and ceramics ($Al_2O_3$)[49,52]. This pronounced disparity between bulk and thin-film stability highlights the importance of kinetic factors and substrate-induced effects in phase selection. In the following, we discuss the differences between α-Ta and β-Ta, the phase selection mechanism in Ta-thin films, and their structural features.

### A. α- and β-phase of Ta

β-Ta crystallizes in a topologically close-packed structure (consisting of icosahedral clusters[53]) and is assigned to the space group $P\overline{4}2_1m$, corresponding to the Frank-Kasper σ-phase structures[54]. In contrast to the majority of σ-phases – which are typically realized in intermetallic systems (with the exception of β-U) – β-Ta constitutes a rare example of a single-element realization of this structural class. There is structure with differentiation of the atoms electron configurations for different positions in the crystal structure[47]. These structural characteristics strongly influence the mechanical response: β-Ta exhibits increased hardness and reduced ductility compared to the α-Ta[55,56]. However, we are more interested in the electrical properties of the various phases of tantalum.

The electrical resistivity of α-Ta thin films typically lies in the range 13 – 30 μΩ×cm[46,50,52,57], comparable to that of moderately conductive metals such as Nb[58,59] and Mo[60,61]. In contrast, β-Ta exhibits substantially higher resistivity, generally exceeding 140 μΩ×cm[50,54,55]. Owing to this pronounced disparity, resistivity serves as a robust diagnostic of phase composition; indeed, even X-ray diffraction (XRD) analysis can be rendered ambiguous by internal stresses in Ta thin films[62]. Films comprising mixed α/β phase fractions display intermediate resistivity values[63–65], consistent with their composite nature. The temperature dependence further differentiates the two phases. For α-Ta, the resistivity decreases upon cooling, as expected for a conventional metal[58,60], and the residual resistance ratio (RRR) exceeds unity. In thin films, RRR values typically remain below ∼20[28,66–68], although significantly larger values – reaching several tens – have been reported[69–73]. By contrast, β-Ta exhibits metalloid behavior[55], with RRR values typically in the range 0.95 – 0.99, indicative of a nearly temperature-independent resistivity[44,50,74]. As an ancillary, though less quantitative, indicator, the two phases can also be distinguished visually: β-Ta generally appears light gray, whereas α-Ta exhibits a darker gray coloration[44,47].

The superconducting critical temperature, $T_C$, is a key parameter for superconducting circuits. For α-Ta thin films, $T_C$ is close to that of bulk Ta ($T_C$ = 4.4 K) [28,44,68,74,75]. In contrast, β-Ta exhibits a markedly reduced $T_C$, with reported values of ≈0.5 K[46], as well as upper bounds of <1.5 K (no transition observed)[50], <2 K (no transition observed)[74], <1.8 K, and <1 K, depending on the fabrication method[76]. Accordingly, α-Ta is the preferred phase for superconducting resonators and qubits. Conversely, β-Ta, owing to its comparatively high resistivity, is well suited for applications in thin-film resistors and heaters[77,78], as well as in spintronic devices exploiting the giant spin Hall effect[79–81].

Beyond superconducting circuit applications, α-Ta thin films are widely employed in a variety of technological contexts. These include their use as diffusion barriers between silicon substrates and copper interconnects[82–84], as electrode materials in electrolytic capacitors[85,86], and as antireflection coatings[87,88]. In microelectromechanical systems (MEMS), α-Ta serves as a structural and functional material for thermal actuators[89,90]. Additionally, its high mechanical robustness and chemical stability make it suitable for wear-resistant coatings in endoprosthetic applications[91,92], as well as for protective coatings in extreme environments, such as gun barrels[93–95]. As noted above, direct deposition of tantalum onto Si, $SiO_2$, and $Al_2O_3$ substrates at room temperature typically results in the formation of the β-phase. Hence, achieving the α-phase requires specific growth strategies. Four principal approaches to stabilizing α-Ta are summarized in Figure 1.

1. **Deposition on a heated substrate.** Reported substrate temperatures for the formation of α-Ta thin films typically fall within the range 400 - 600 °C[44,75,96–98], with a minimum value of approximately 350 °C[63,99]. A separate set of studies has explored substantially higher temperatures in the range 900 - 1550 °C [73]. These observations suggest the existence of a threshold temperature for stabilizing the α-phase; however, its precise determination remains challenging. In particular, the presence of oxygen – either in the residual atmosphere, the working gas, or at the substrate surface – tends to increase the temperature required for α-phase formation[62]. Moreover, deposition onto an insufficiently heated substrate often results in mixed α+β phase films[44,63,67], where a minor β-phase fraction may remain undetected by standard characterization techniques. Additional uncertainty arises from inaccuracies in the measurement of the actual substrate temperature during growth. A recent study[100] reported the formation of α-Ta on Si substrates at temperatures as low as 200 °C by employing krypton as the working gas in place of argon, highlighting the sensitivity of phase stabilization to deposition conditions. Further study is required.

2. **Deposition on a buffer layer.** The stabilization of α-Ta at room substrate temperature can be achieved via the introduction of a metallic buffer layer, typically 5-10 nm thick, provided that the buffer remains free of oxidation. Under high-vacuum conditions and within a single deposition cycle, α-Ta growth has been demonstrated on buffer layers composed of Nb[44,67,74,99], Al[44,51], Mo[44,63], Ti[51,101,102], Cr[103,104], and W[49,105], as well as on noble metals such as Pt[49,51], Au[49,51], Ru[101], and Rh[49]. In contrast, exposure of the buffer layer to air prior to Ta deposition, leading to surface oxidation, promotes the formation of the β-phase. This behavior has been observed for oxidized Nb[44], Al[44,51], Ni[49], Cu[49], and Ti[106] buffer layers. For non-metallic buffer layers, the phase outcome at room temperature depends sensitively on the specific material: α-Ta growth has been reported on TaN[76,106–108] and TiN[76,109,110], whereas β-Ta is typically obtained on amorphous $SiN_X$[63] and AlN[76].

3. **Annealing of the β-Ta film.** β-Ta is a metastable phase that transforms into the thermodynamically stable α-phase upon thermal annealing. The required annealing temperature typically exceeds the substrate temperatures employed during direct α-phase deposition, ranging from approximately 600°C[111] to 750°C[64,103] for annealing processes involving a break in vacuum. The presence of tantalum oxide is believed to stabilize the β-phase, thereby increasing the temperature required for the phase transition. Consequently, higher vacuum levels during annealing, corresponding to reduced oxygen partial pressure in the residual atmosphere, facilitate the transformation and lower the required annealing temperature[111,112].

4. **Deposition at cryogenic temperatures.** It has been shown that deposition of Ta on various substrates (Si, $Al_2O_3$, GaAs, $SiN_X$) at temperatures below 50 K ensures the growth of the α-phase[68]. It is assumed that at cryogenic temperatures, the Ta-amorphous phase grows, which, upon heating the film to room temperature, transforms into stable α-Ta, bypassing the metastable β-phase.

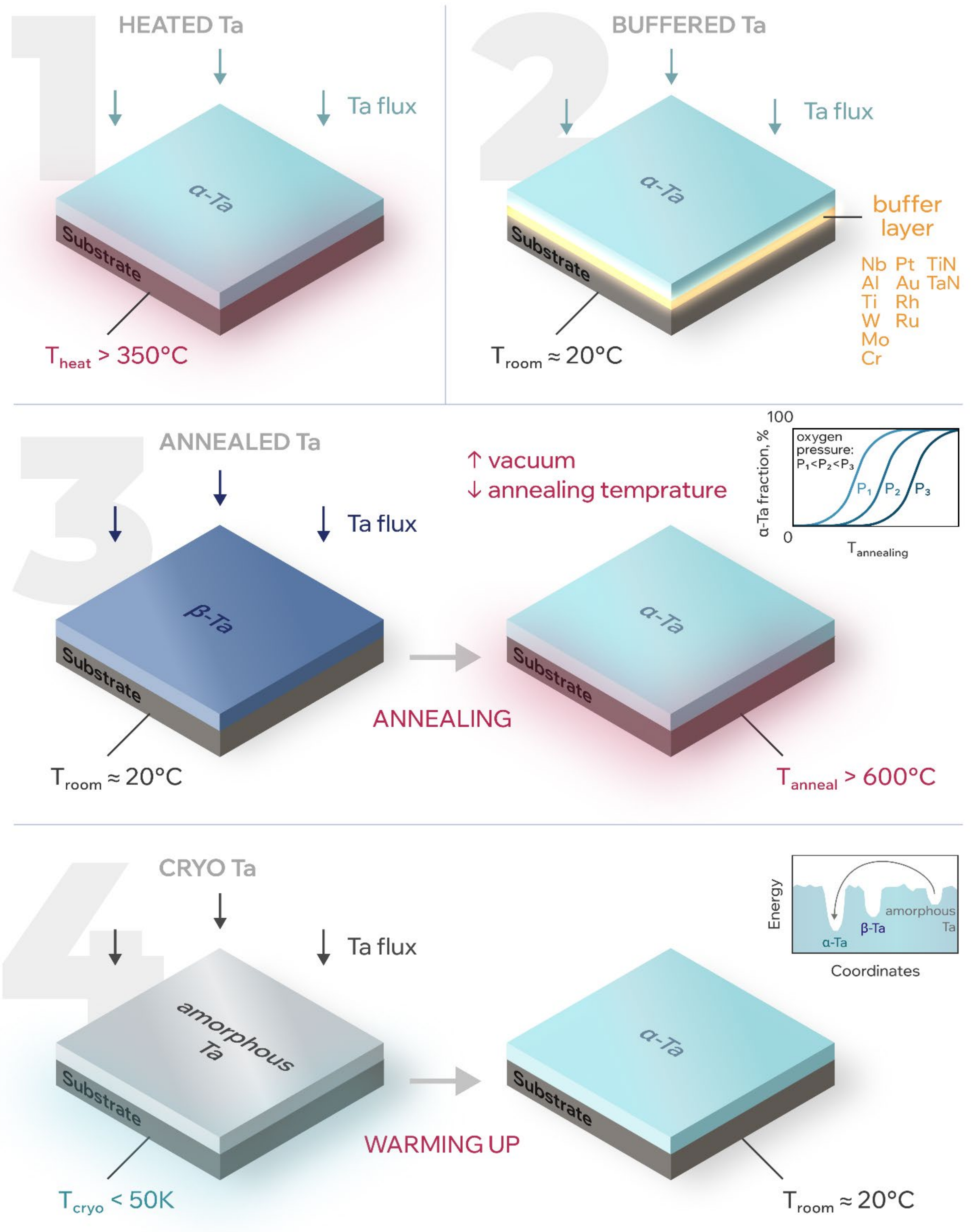

*Figure 1*

Approaches to forming thin tantalum films in the alpha-phase. The insert for ANNEALED Ta schematically shows the dependence of the alpha-phase fraction in the film on the annealing temperature at various residual oxygen pressures. The insert for CRYO Ta schematically shows the potential energy landscape for various tantalum phases and the transition from the amorphous phase to the alpha-phase, bypassing the beta-phase, during cryogenic deposition.

### B. The phase selection mechanism in Ta films

The mechanism of phase selection in Ta thin films remains an open question; however, a substantial body of experimental evidence enables several qualitative conclusions. As discussed in Sec. II.A, the presence of oxygen in the residual atmosphere during deposition on a heated substrate, or during post-deposition annealing of β-Ta, necessitates higher substrate or annealing temperatures to achieve complete transformation to the α-phase. Similarly, oxidation of a metallic buffer layer suppresses the nucleation of α-Ta. These observations collectively indicate that oxygen plays a key role in stabilizing the β-phase. In this context, a mechanism has been proposed[62] based on the epitaxial growth of β-Ta on a δ-$TaO_2$ interfacial oxide layer that forms at the initial stages of film deposition. Within this framework, the requirement for elevated substrate temperatures is attributed to the desorption of oxygen and water from the substrate surface, thereby reducing oxide formation and promoting α-phase nucleation.

Experiments on the growth of Ta films at different temperatures on a-$SiN_X$ buffer layer suggested[63] that a transition layer of amorphous Ta forms on amorphous buffer layers (which also include native metal and silicon oxides), over which either the α-phase or β-phase grows, depending on the substrate temperature. Temperature in this case determines the energy landscape of the substrate surface, and up to a certain temperature (≈ 600 K), the energy barrier to nucleation for β-Ta is lower than for α-Ta. In the case of metallic buffer layers (freshly deposited Mo), α-Ta grows according to a pseudomorphic-epitaxial mechanism.

An earlier series of experiments on Ta deposition onto a range of substrates likewise demonstrated[51] that the use of buffer layers with an ordered crystalline structure (e.g., as-deposited Ti) promotes the growth of α-Ta. In this work, the authors also argued that the β-phase is energetically favored at low temperatures. However, they proposed a different interpretation of the role of oxygen at the substrate surface. Specifically, interdiffusion between the substrate (or buffer layer) and the deposited tantalum was found to favor α-phase formation. In contrast, the presence of surface oxides suppresses such interdiffusion, thereby stabilizing the β-phase.

Thus, whatever the actual mechanism, all researchers agree that the surface oxide layer, adsorbed oxygen, and oxygen in the residual atmosphere are responsible for the stabilization of β-Ta. In a recent study[113], density functional theory (DFT) modeling showed that when Ta is doped with oxygen above ≈ 13 at.%, the tetragonal structure of β-Ta is energetically more favorable. However, the question remains: why has not the reliable production of α-Ta on the extremely common silicon substrates been demonstrated without heating? In a recent study[44], an attempt was made to produce α-Ta on Si at room temperature using preliminary surface cleaning from oxide and adsorbates by treatment in a hydrofluoric acid (HF) solution, argon plasma etching, and high-temperature annealing without breaking the vacuum cycle. None of this ensured the growth of the α-phase. Moreover, although the β-phase grew on the Nb and Al buffer layers oxidized in air, argon plasma etching, by removing the surface oxide layer, allowed growth of α-Ta. Thus, the effect of substrate heating is certainly not limited to the removal of adsorbates from the surface. This still does not contradict the assumption[63] of a lower nucleation energy barrier for β-Ta compared to α-Ta at room temperature, but in this case, α-Ta would grow on $Al_2O_3$ substrates at room temperature after plasma etching or high-temperature annealing due to the low lattice mismatch between the c-plane sapphire and α-Ta (111) (less than 1.7%). However, this does not occur.

Thus, irrespective of the specific microscopic mechanism, there is broad consensus that surface oxides, adsorbed oxygen, and oxygen present in the residual atmosphere play a decisive role in stabilizing the β-phase of Ta. Consistent with this view, a recent density functional theory (DFT) study[113] demonstrated that oxygen incorporation above ∼13 at.% renders the tetragonal β-Ta structure energetically favorable. Despite this progress, a key question remains unresolved: why has the reproducible formation of α-Ta on technologically ubiquitous Si substrates not been achieved at room temperature? In Ref.[44], extensive surface preparation protocols – including oxide removal via hydrofluoric acid (HF) treatment, argon plasma etching, and high-temperature annealing without breaking vacuum – were employed prior to deposition. Nevertheless, these measures did not result in α-phase growth. At the same time, although β-Ta forms on Nb and Al buffer layers exposed to air, subsequent argon plasma etching – which removes the surface oxide – enables α-Ta growth. This observation indicates that the role of substrate heating extends beyond the mere desorption of surface contaminants. While these findings remain consistent with the hypothesis[63] that β-Ta possesses a lower nucleation barrier than α-Ta at room temperature, they also expose limitations of a purely thermodynamic or kinetic description. For example, one might expect that, following oxide removal (e.g., by plasma etching or high-temperature annealing), α-Ta would nucleate on $Al_2O_3$ substrates at room temperature, given the small lattice mismatch (less than 1.7%) between c-plane sapphire and α-Ta (111). Experimentally, however, this is not observed, underscoring that additional factors govern phase selection in Ta thin film growth.

To explain the absence of α-Ta growth on Si and $Al_2O_3$ without heating, it was suggested[44] that heating is the energetic requirement for the formation of the α-phase of Ta on clean substrates such as Si and $Al_2O_3$. The authors noted that, according to the results of molecular modeling[53,114], tantalum has a low cooling rate from the melt ($10^{11}$–$10^{13}$ K/sec) for obtaining amorphous metallic glass. As mentioned in Section II.A, β-Ta has a topologically close-packed structure consisting of icosahedral clusters similar to those that form Ta-metallic glass[115]. Thus, β-Ta grows at fairly high crystallization rates, but lower than those required to form amorphous Ta[53]. What determines the crystallization rate during thin film growth? Atoms deposited on a substrate lose excess energy over several lattice vibration periods of the substrate material[116,117]. The maximum lattice vibration frequency is determined by the Debye temperature of the substrate ($\theta_D$)[118,119]. The authors compared[44] the $\theta_D$ of the substrate material and the tantalum phase formed at room temperature and found that β-Ta forms on substrates with higher $\theta_D$. It was proposed that high $\theta_D$ of the substrate material ensures rapid cooling of the deposited atoms, leading to the growth of the Ta β-phase. When the substrate is heated, the temperature difference between the deposited atom and the substrate decreases, resulting in a decrease in the cooling rate, leading to the crystallization of α-Ta. In addition, oxides usually have a higher $\theta_D$[120,121], so α-Ta grows on freshly prepared buffer layers of Nb and Al type, but β-Ta already grows on the same materials subjected to oxidation.

To account for the absence of α-Ta growth on Si and $Al_2O_3$ substrates without heating, it has been proposed[44] that thermal activation is intrinsically required for α-phase formation on otherwise clean surfaces. The authors noted that, according to the results of molecular modeling[53,114], tantalum has a low cooling rate from the melt ($10^{11}$-$10^{13}$ K/sec) for obtaining amorphous metallic glass. As mentioned in Section II.A, β-Ta has a topologically close-packed structure consisting of icosahedral clusters similar to those that form Ta-metallic glass[115]. This structural similarity suggests that β-Ta forms under relatively high crystallization rates, albeit lower than those necessary for complete amorphization[53]. A key question, therefore, is what governs the effective cooling (crystallization) rate during thin-film growth. Atoms deposited on a substrate lose excess energy over several lattice vibration periods of the substrate material[116,117]. The maximum lattice vibration frequency is determined by the Debye temperature of the substrate ($\theta_D$)[118,119]. In Ref. [44], a correlation was identified between $\theta_D$ and the phase formed at room temperature: β-Ta tends to grow on substrates with higher Debye temperatures. Within this model, a high $\theta_D$ facilitates rapid energy dissipation from the deposited atoms, corresponding to a high effective cooling rate that favors β-phase formation. When the substrate is heated, the temperature difference between the deposited atom and the substrate decreases, resulting in a decrease in the cooling rate, leading to the crystallization of α-Ta. In addition, oxides usually have a higher $\theta_D$[120,121], so α-Ta grows on freshly prepared buffer layers of Nb and Al type, but β-Ta already grows on the same materials subjected to oxidation.

Thus, the hypothesis invoking the influence of the substrate Debye temperature provides a consistent framework for interpreting a large fraction of the experimental observations. However, an important discrepancy remains: the minimum substrate temperature required for α-Ta formation is nearly identical for Si and $Al_2O_3$, despite their substantially different Debye temperatures ($\theta_D$ is 650 K[122] for Si and 1000 K[123] for $Al_2O_3$). This observation suggests that the applied substrate heating may be insufficient to significantly reduce the instantaneous temperature difference between incoming adatoms and the substrate, and hence does not fully suppress the high effective cooling rates associated with β-phase nucleation. Instead, the heating appears sufficient to induce a rapid transformation of the initially formed β-phase into the thermodynamically stable α-phase during growth. Within this interpretation, deposition onto a heated substrate can be viewed as a form of in situ “flash annealing,” wherein the film undergoes continuous, thermally activated phase relaxation concurrent with deposition.

It is worth noting that at least one study[56] reports the formation of α-Ta on a Si substrate at room temperature and without the use of a buffer layer, but only within a narrow range of sputtering pressures. However, a subsequent study[62] did not reproduce this behavior and found no evidence of such a distinct technological regime over a wide range of working pressures. These inconsistencies highlight the limited reproducibility and incomplete understanding of phase selection in Ta thin films. Clearly, theoretical and experimental studies of the phase selection mechanism in Ta thin films need to be continued.

### C. Structure of α-Ta films on Si and $Al_2O_3$ substrates

Since Si and $Al_2O_3$ are key substrate materials for superconducting circuits, we focus on the structural properties of α-Ta films grown directly on these substrates and on intermediate buffer layers. Since α-Ta has a BCC lattice, its films can grow according to three orientations: (100), (110), and (111). These orientations will correspond to figures in the cross section perpendicular to the growth direction: a square with the parameter a = 3.303 Å for (100), a pseudohexagon with the parameters $a_1$ = 2.86 Å and $a_2$ = 3.31 Å for (110)[108], and a hexagon with the parameter a = 4.68 Å for (111). Table 1 lists potential substrates (buffer layers) for growing α-Ta, indicating their orientation, cross-sectional shape, and parameters, as well as the expected corresponding α-Ta orientation and parameter mismatch (for α-Ta, (110) corresponds to 2 parameters). For buffer layers with a BCC structure and small lattice mismatch, α-Ta is generally expected to adopt the same crystallographic orientation via epitaxial alignment. Table 1 shows that the Si (100) lattice is completely mismatched with the α-Ta lattice, so in this case, only polycrystalline α-Ta films are expected to grow. In the case of the $Al_2O_3$ c-plane, pseudomorphic epitaxial growth of α-Ta with the (111) orientation is expected, and the closest lattice match for α-Ta is observed with Nb. This is one of the reasons why Nb is most often used as a buffer layer for the formation of α-Ta without heating. In addition, Ta and Nb are natural neighbors, meaning they are found together in mineral ores[124,125]. Therefore, in Ta-based materials for thin film deposition, Nb is the largest impurity[126].

*Table 1*

Substrate material (Substr.) comparison indicating the crystal lattice type (Lattice), crystal orientation (Orient.), substrate surface figure (Figure), lattice parameter(s) (Param, Å), and corresponding α-Ta film orientations (Orient. α-Ta) with lattice parameter mismatch as a percentage (Mismatch 1 and 2, if applicable). "Any" in the orientation column indicates that the α-Ta orientation mirrors the substrate material orientation.

| Substr. | Lattice | Orient. | Figure | Param, Å | Orient.α-Ta | Mismatch 1, % | Mismatch 2, % |
|---|---|---|---|---|---|---|---|
| TaN | FCC | 111 | hexagon | 3.07 | 110 | 6.84 | 7.25 |
| Ti | hexagonal | 100 | hexagon | 2.9505 | 110 | 3.07 | 12.18 |
| W | BCC | 100 (any) | square | 3.16 | 100 (any) | 4.75 | - |
| Nb | BCC | 100 (any) | square | 3.3 | 100 (any) | 0.3 | - |
| Mo | BCC | 100 (any) | square | 3.14 | 100 (any) | 5.41 | - |
| Al | FCC | 111 | hexagon | 2.863 | 110 | 0.1 | 15.61 |
| Cu | FCC | 111 | hexagon | 2.556 | - | 11.89 | 29.50 |
| Cr | BCC | 100 (any) | square | 2.88 | 100 (any) | 14.93 | - |
| $Al_2O_3$ | hexagonal | c-plane | hexagon | 4.758 | 111 | 1.62 | - |
| $Al_2O_3$ | hexagonal | a-plane | rectangle | 4.33<br>4.12 | 100 | 23.72 | 19.83 |
| Si | FCC | 100 | square | 5.431 | 100 | 39.05 | - |

Traditionally, XRD is used to determine the orientation of α-Ta films; however, studies[44,99] demonstrate the high information content of the electron backscatter diffraction (EBSD) method for this task. Experiments show that α-Ta films typically grow with the (110) orientation on both Si[66,68,75,76,99] and $Al_2O_3$[66–68,96,97], as well as on crystalline Nb[50,67,74,99,107] and amorphous $TaN_X$[76,107,108] and $TiN_X$[110] (when deposited on a (111) Nb buffer layer with heating, α-Ta (111) growth is possible[72]). In this case, α-Ta films growing on buffer layers copy the morphology of the buffer layer[76,127], and on different orientations of Si[66] and $Al_2O_3$[41,67,128] the morphology of α-Ta (110) is always the same, and the film surface is a set of elongated, closely spaced grains (Figure 2) according to the results of scanning electron microscopy (SEM) and atomic force microscopy (AFM). Moreover, on Si, the film surface additionally contains smooth regions[43,45,71] (Figure 2), in addition to regions with elongated codirectional grains that form ridges (the only mention[129] of a similar structure for $Al_2O_3$). In low-magnification images (Figure 2), it is evident[44,71,99] that the film consists of large cells with a size of tens of micrometers, and the ridges form a centripetal pattern. Studies[44,130] have shown that the first approximately 10 nm of Ta on a heated Si substrate grow in the β-phase, and α-Ta regions nucleate pointwise on this underlayer, rapidly expanding with increasing film thickness. The β-Ta underlayer is presumably stabilized by the interdiffusion of Si and Ta[112,131,132].

Pseudomorphic epitaxial growth of α-Ta with (111) orientation was achieved only on $Al_2O_3$ with c-plane orientation[70–72,134] (Figure 2), while only α-Ta (110) grew on $Al_2O_3$ a-plane[97,128,133,135]. The epitaxial nature of α-Ta (111) growth is confirmed by the record RRR values of these films: more than 40[70], more than 60[71] and even more than 70[72]. In addition, it was shown[72] that upon damaging the $Al_2O_3$ substrate surface with argon ions, α-Ta with (110) orientation grows instead of (111) with a simultaneous radical deterioration of RRR from 45 to 10 and root-mean-square roughness (Rq) from 3.5 to 13.3 Å. A similar deterioration in RRR and Rq was observed in[70] upon transition from the (111) to (110) orientation as a result of changing the deposition mode. Thus, the preferred orientation for polycrystalline α-Ta film growth is (110), and this orientation will therefore be formed when using lattice-mismatched crystalline substrates (primarily Si) or amorphous substrates (buffer layers). Pseudomorphic epitaxial growth of α-Ta (111) can be achieved on $Al_2O_3$ c-plane under certain process parameters.

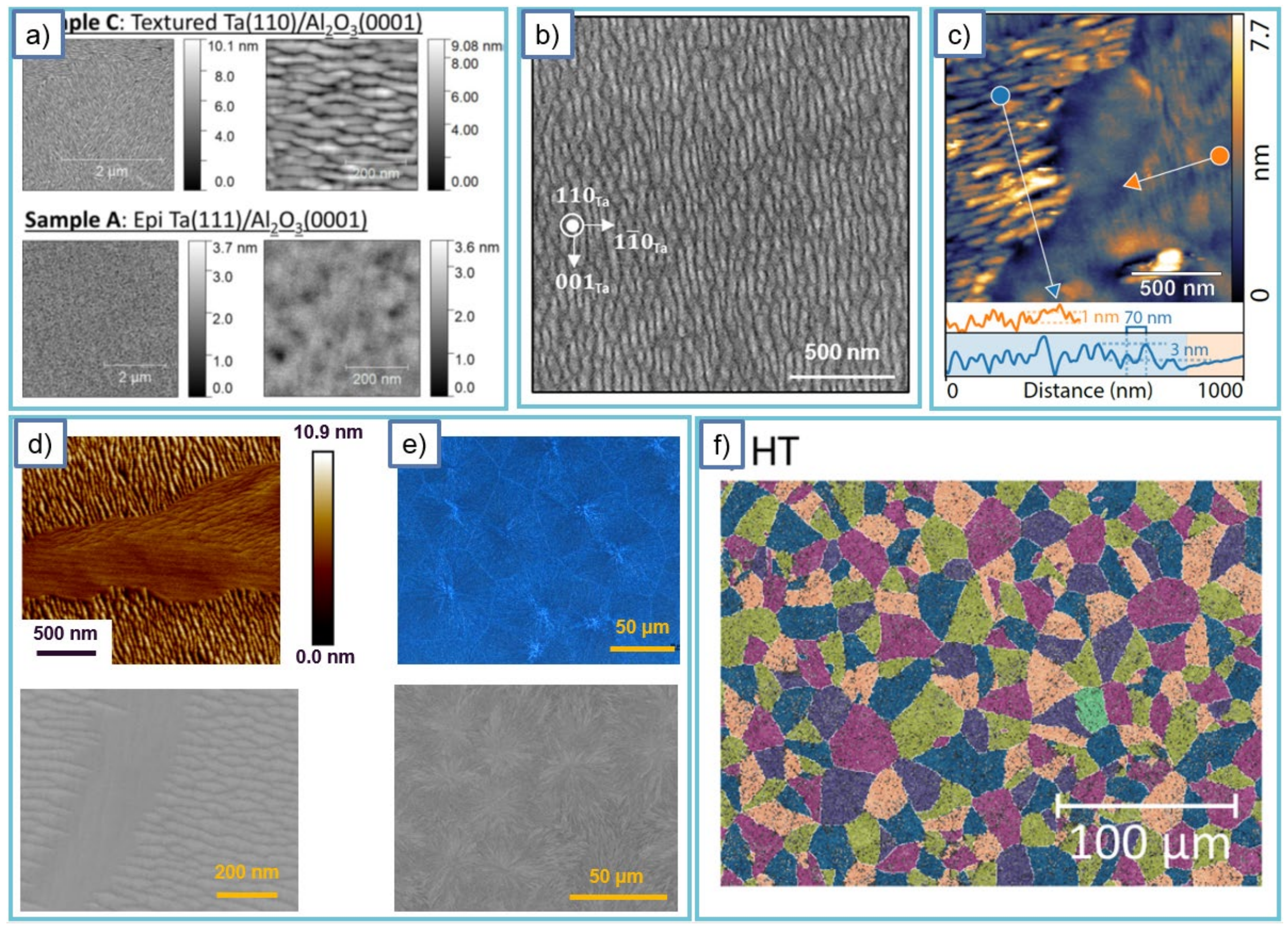

*Figure 2*

Morphology of α-Ta thin films. a) AFM images of α-Ta (110) and (111) on the c-plane of $Al_2O_3$, reprinted from[72], Copyright (2025) American Physical Society . b) SEM image of α-Ta (110) on the a-plane of $Al_2O_3$, reprinted from[133], Copyright (2023) American Chemical Society. c) AFM image of rough and smooth regions of α-Ta (110) on the c-plane of $Al_2O_3$ (the only evidence of mixed morphology on sapphire) reprinted from[129], Copyright (2023) American Chemical Society. d) AFM and SEM images of rough and smooth regions of α-Ta (110) on Si (100). e) Optical dark-field and SEM images of the coarse-mesh structure of α-Ta (110) on Si (100) reprinted from[44], Copyright (2026) American Institute of Physics. f) False-color SEM image of the coarse-meshed α-Ta (110) structure on Si (100), reflecting the crystal orientation determined by EBSD, reprinted from[99], under a Creative Commons license (https://creativecommons.org/licenses/by/4.0/).

## III. Fabrication of α-Ta resonators and qubits

The fabrication process of α-Ta-based superconducting circuits (Figure 3) is similar to that used for aluminum[21–23]. However, due to its chemical resistance and high melting point, Ta circuits are compatible with various post-treatment techniques, including acid treatment and high-temperature annealing. This is likely one of the key reasons for the superior performance of α-Ta circuits[41–45,75,99]. Below, we discuss the steps involved in the formation of the Ta-base layer.

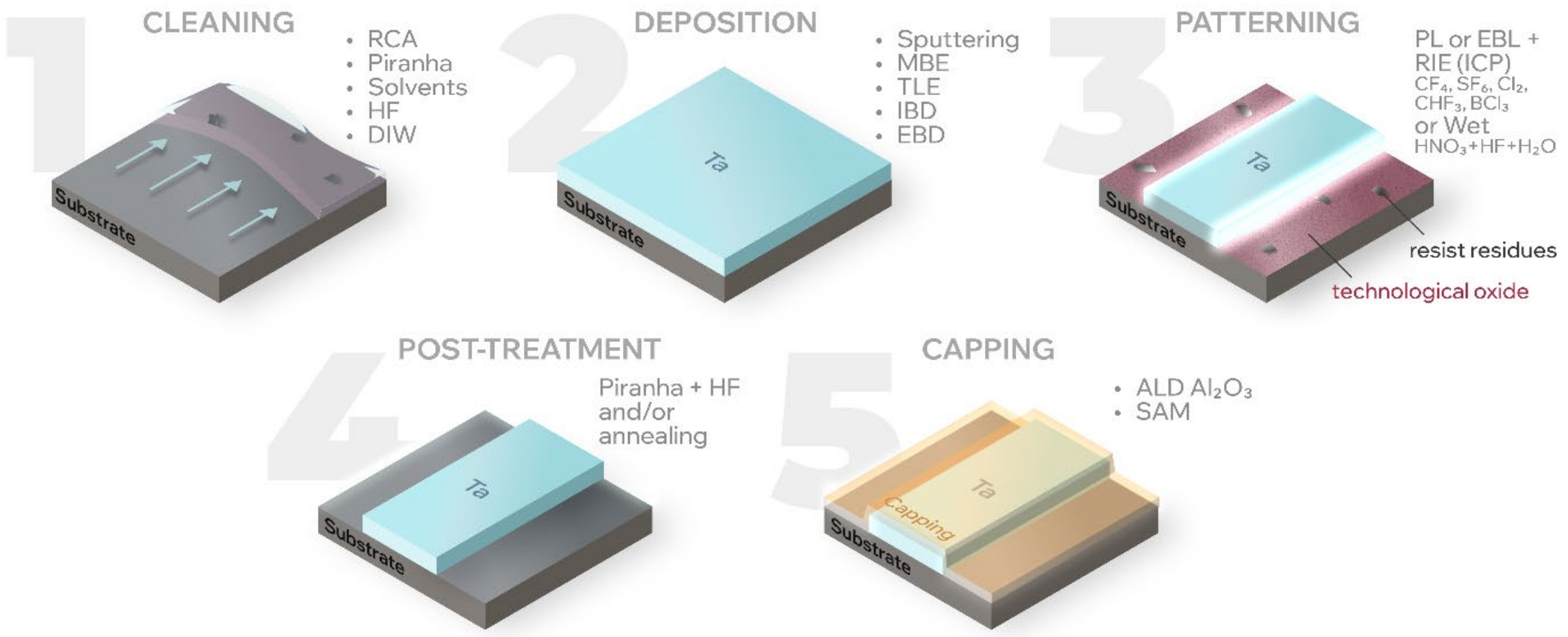

*Figure 3*

The base layer fabrication process of a superconducting qubit or resonator based on alpha-tantalum. MBE: Molecular Beam Epitaxy. TLE: Thermal Laser Epitaxy. IBD: Ion Beam Deposition. EBD: Electron Beam Deposition. PL: Photolithography. EBL: Electron Beam Lithography. RIE: Reactive Ion Etching. ICP: etching with Inductively Coupled Plasma. Wet: Wet etching. ALD: Atomic Layer Deposition. SAM: Self-Assembled Monolayer capping.

### A. Substrate preparation and α-Ta film deposition

RCA-cleaning of Si wafers is considered the standard in semiconductor manufacturing; however, likely due to its complexity, only a few reports of its use in Ta circuit fabrication exist[107,136]. More commonly, organic solvents[76,113,137] (acetone, isopropyl alcohol) or Piranha solution[43,44,68] ($H_2SO_4$ and $H_2O_2$) are used to remove organic contaminants, together with hydrofluoric acid solutions[66,67,74,75,99] (HF or buffered oxide etchant, BOE) to remove the native silicon oxide and suppress reoxidation for several hours. Removal of organic contamination is particularly important because Ta deposition is often performed on square wafers obtained by cutting or cleaving standard round wafers coated with a protective resist. Additionally, For Si (111), a 7×7 surface reconstruction can be achieved[68] by annealing in a hydrogen atmosphere at 800 °C, followed by vacuum annealing at 1000 °C.

$Al_2O_3$ wafers are likewise cut using a resist; accordingly, cleaning in Piranha solution[66,68,70,134] and organic solvents[69,72,96] is commonly employed. Ozone treatment[66,127,138] is also used. In addition, sapphire substrates are frequently annealed in situ in the deposition chamber at temperatures exceeding 800 ∘ C to remove surface contamination[72,97,134,135,139]. Both Si and $Al_2O_3$ substrates may also be cleaned by ion bombardment in the deposition chamber[69,74]; however, this can induce β-Ta inclusions[69] or alter the preferred orientation of α-Ta[72] due to surface damage associated with ion irradiation.

Thin Ta films can be deposited by various methods: electron beam evaporation[103,130,131,140], molecular beam epitaxy (MBE)[68,97,141], thermal laser epitaxy[73], ion beam deposition[104,142], but the most common method is magnetron sputtering. Most researchers use DC-sputtering[66,75,98,106,134,143], but RF-sputtering[76,133] is also used, probably due to the lack of a DC-power source. Taking into account the information from Section II.A., we note that for the formation of α-Ta films, the sputtering system must be capable of heating the substrate to 500 °C or additional sputtering sources for depositing a buffer layer without breaking the vacuum. Traditionally, the film thickness of α-Ta for the base layer of superconducting circuits is 100[75,99,137], 150[44,107,127] or 200 nm[41,134,141], however, both thinner films[68,135,139] and thicker ones[71,144] can be successfully used.

The formation of α-Ta in the absence of a buffer layer typically requires substrate heating to approximately 500 °C, while modern thin-film deposition systems enable substrate temperatures exceeding 1000 °C. It has been shown[44] that heating Si substrates above 600 °C leads to a pronounced increase in surface roughness. However, as reported in Ref.[73], the roughness reaches a maximum near 900 °C and subsequently decreases up to 1150 °C, where large, flat, densely packed grains are observed. At still higher temperatures, the film undergoes morphological instability and separates into individual grains. These observations suggest that, for α-Ta deposition with substrate heating, it is advantageous either to limit the temperature to ~600 °C or to operate in the high-temperature regime above 1000 °C. Notably, a recent report[100] demonstrated the formation of a pure α-phase at substrate temperatures as low as 200 °C using magnetron sputtering in a krypton atmosphere, underscoring the importance of exploring alternative working gases in Ta film growth, particularly xenon[95].

### B. Patterning of α-Ta film

The formation of the α-Ta base layer in superconducting circuits is typically carried out using a standard subtractive process, involving the definition of a resist mask followed by film etching. The resist pattern is most commonly defined by direct laser writing[41,42,44,134], although electron-beam lithography[69,75,145] is also employed. Etching of Ta can be performed using wet chemical methods[41,97,134,146], for example in mixtures of HF and $HNO_3$ and water, as well as by dry reactive ion etching (RIE) in fluorine-based chemistries ($CF_4$, $SF_6$, $CHF_3$, $CH_2F_2$)[66,67,76] or chlorine-based chemistries ($Cl_2$, $BCl_3$, $SiCl_4$) [68,70,134,147], with or without the use of inductively coupled plasma (ICP).

Wet etching does not require specialized equipment, and commercially available etchants are readily available. However, careful process design[41] or encapsulation of the α-Ta film, for example with AuPd[146], is required due to possible underetching beneath the resist[41,148] associated with film delamination. RIE provides a more productive and reproducible approach[42], but likewise demands careful optimization to avoid etching defects[41]. A key advantage of RIE is the ability to achieve deep anisotropic etching of Si substrates, enabling the fabrication of trench α-Ta circuits[43,44]. However, this capability is accompanied by significant substrate damage: SEM images indicate that RIE of α-Ta on Si without a buffer layer leads to pronounced surface degradation[43,44], which may adversely affect subsequent Josephson junction fabrication[22,149]. Moreover, even chemically robust $Al_2O_3$ substrates are susceptible to damage under RIE conditions due to ion bombardment[146]. It should also be noted that, because oxidation of the Ta surface is inherent to the wet etching process, the resulting oxide layer is approximately twice as thick as that formed after RIE[146]. A direct comparison of device performance for Ta circuits fabricated using dry and wet etching would therefore be valuable, as previously demonstrated for Al-based systems[21,150]. Overall, dry etching represents a preferred approach for patterning high-performance α-Ta circuits; however, precise control of process parameters is essential to minimize substrate damage.

### C. Post-treatment and capping of α-Ta circuits

As noted above, the compatibility with post-processing treatments is a key advantage of α-Ta circuits compared to Nb and, especially, Al-based circuits. In the earliest demonstrations of Ta qubits with record relaxation times[41,42], post-treatment in Piranha solution was applied after each solvent-based resist stripping step to remove residual organic contamination. In addition, acid treatment is reported to produce a more uniform tantalum oxide with reduced suboxide content and, consequently, a lower density of two-level systems (TLS), despite the increased oxide thickness compared to the native oxide[42,134,151]. Piranha post-treatment is currently employed for both Si-based[43,44] and sapphire-based devices[71,134,144,151]. No detrimental effects of this treatment on superconducting circuit performance have been reported to date. At the same time, systematic studies unambiguously demonstrating its effectiveness remain lacking.

The post-treatment of Ta circuits in HF solutions, such as 2%[74] or 10%[45,75] aqueous HF and BOE 6:1[66,68], 7:1[69,76] or 10:1[43,99,134] buffer etchant, has obvious and confirmed effectiveness. The treatment time varies greatly from 1 or 2 minutes[68,72,75] to 15 or 20 minutes[43,69,134]. The key to increasing the figure of merit of Ta circuits is the removal of silicon oxides in the base layer gaps and tantalum oxide from the base layer surface[44,134], as is the case with Nb circuits[152]. Due to the passivation of the silicon surface in HF, the effect of the post-treatment lasts for several hours, during which the growth of natural $SiO_2$ is significantly suppressed[45,75,153]. As a result, a record-breaking internal Q-factor of over 10 million was demonstrated for Ta-on-Si resonators at low power[43,44]. Unfortunately, due to the longer fabrication process, it has not yet been possible to reproduce this effect in qubit circuits. Since the etching rate of $Ta_2O_5$ in HF solutions is significantly lower than that of $SiO_2$ and, after treatment, $SiO_2$ is completely removed, while $Ta_2O_5$ is only partially removed. It can be assumed that the effect of decreasing the $Ta_2O_5$ thickness on the circuit quality is much smaller than the effect of decreasing the $SiO_2$ thickness[45,74]. Furthermore, it has been demonstrated[45] that prolonged exposure of Ta circuits to HF solution can lead to a significant decrease in the circuit quality due to the formation of non-superconducting tantalum hydride and even damage to the base layer.

Another method for post-treatment of Ta circuits was proposed in a study[45]: high-temperature annealing of already formed circuits at 500°C for 1 hour. This is currently the only example of achieving a Q-factor of Ta resonators at low power, exceeding 10 million, without the use of HF post-treatment. A recent study[154] proposed an approach to modifying the surface oxide of Ta by deposition a 0.2-nm-thick sacrificial titanium layer onto the α-Ta film, which is then removed in a BOE prior to cryogenic characterization. Additionally, post-annealing of the Ta circuits at 700 °C was performed, resulting in a 4-fold increase in resonator performance. Thus, the best solution for post-treatment Ta circuits is currently a combination of Piranha solution and HF solution, but new approaches need to be explored.

Capping (encapsulation) of the base layer surface to prevent native oxide growth can be an effective way to improve the performance of superconducting circuits[155,156]. Capping is typically performed immediately after base layer film deposition, i.e., before patterning, leaving the sidewalls unprotected, but the electric field concentration is highest in the sidewall region[157]. Incidentally, Ta has shown high efficiency as an encapsulation layer for Nb[136]. Sputtered amorphous 2 nm $Al_2O_3$ was used to protect the Ta-base layer, maintaining the Q-factor of the resonators for over 1 year, while without encapsulation, the Q-factor significantly decreased after just 2 months[135]. Au and AuPd coatings with a thickness of 4–5 nm have also been shown to be effective in suppressing $Ta_2O_5$ growth on the top surface of the base layer[157]. It is important to note that incomplete coverage (island film) of the Ta surface by the AuPd layer can lead to increased surface oxidation due to the catalytic effect[157]. However, in the cited studies, suppression of native tantalum oxide growth did not result to a significant increase in circuit performance. This is likely due to the fact that native tantalum oxide contains a small amount of potential TLS sources[133,136,158,159]. Therefore, at present, it is advisable to focus efforts on methods of passivation and encapsulation of the silicon surface in the base layer gaps, for example, using SAM (self-assembled monolayers)[160,161] or thin ALD (atomic layer deposition) films[162], and not only on encapsulation of the top Ta surface.

## IV. Performance of α-Ta resonators and qubits

The performance of superconducting circuits is determined by their ability to preserve an excited state with minimal dissipation and decoherence. Because qubits operate at low signal powers, a key figure of merit for resonators is the internal quality factor in the single-photon regime, $Qi_{LP}$. This quantity is typically several times lower than the high-power quality factor, $Qi_{HP}$, due to enhanced energy dissipation arising from parasitic two-level systems (TLS) at low excitation powers[20,163]. The choice of material and fabrication technology for the base layer plays a central role in determining the TLS density in the circuit. For qubits, the corresponding figure of merit is the energy relaxation time $T_1$, which depends on multiple factors, including TLS contributions associated with the base layer. In the following, we review the performance achieved in tantalum-based resonators and qubits, analyze the factors underlying the advantages of Ta circuits, and discuss the remaining open questions in Ta-based superconducting devices.

### A. Characterization of superconducting resonators and qubits

Detailed information on the design and characterization of superconducting resonators can be found in a review[163]. Here, we present only the essential concepts.

The most common superconducting resonator in quantum circuits is a λ/4 (less commonly λ/2) coplanar waveguide resonator (CPWR), i.e., an LC-resonator with inductance and capacitance distributed along its length. Lumped-element (LE) resonators are also used in some cases[66,69,134]. The key geometric parameters of a CPWR are its length, the width of the central conductor, w, and the gap between the conductor and the ground plane, g. Typically, a chip with resonators includes a single 50-Ω feedline with two contacts, to which 2 - 20 resonators with different resonant frequencies are capacitively or inductively coupled. The coupling element between the resonator and the feedline (the coupler) is usually arranged parallel to the feedline. Increasing the length of the coupler enhances the coupling strength, corresponding to a reduction in the coupling quality factor $Q_C$. In some designs, to equalize the potential of the two halves of the ground plane separated by the feedline, air bridges[21,24,164] are fabricated above the feedline or the connection is implemented using wire bonds[43,74].

The resonator characterization procedure includes measuring the amplitude-frequency response of $S_{21}$ in the frequency range of 200–500 kHz around the resonance of each resonator using a vector network analyzer (VNA). The complex response of $S_{21}$ is scanned from low to high powers, and the resonant frequency, loaded quality factor Ql, Qc, and internal quality factor Qi (as a function of power) are extracted, for example, using the open-source package described in detail in work[165]. The power limits are chosen such that low power corresponds to the single-photon regime in which $Qi_{LP}$ is measured, while high power does not drive the resonator into a nonlinear regime. Measurements are performed when the resonator is in the superconducting state, and the temperature is typically no higher than 50 mK to minimize thermal noise, with the sample carefully shielded from magnetic and infrared noise[166,167]. The input signal from the VNA is attenuated using attenuators, and the output signal is traditionally amplified using a parametric amplifier[14,15] (optional) and high-electron-mobility transistor (HEMT) amplifiers[21].

An important geometric parameter of a resonator is its surface participation ratio (SPR). This is an integral parameter proportional to the fraction of the electromagnetic field confined within the circuit interfaces[168]. SPR is the sum of the participation ratios of three interfaces: metal-substrate $p_{MS}$, metal-air $p_{MA}$, and substrate-air $p_{SA}$. The higher the SPR, the more strongly TLS contribute to the reduction of $Qi_{LP}$ of the resonator. The main ways to reduce SPR are to increase the gap g and to introduce trenches in the gap between the resonator center conductor and the ground plane[32,169]. When comparing the $Qi_{LP}$ of resonators with different geometries, the corresponding differences in SPR must be taken into account. LE resonators typically exhibit lower SPR values than CPWRs.

Random uncontrolled processes in the qubit environment, including microscopic contaminants and defects on the qubit chip, chip packaging, control and measurement electronics, act as noise sources that lead to qubit decoherence and thus a decrease in gate fidelity.

In a typical system, the qubit is weakly coupled to these noise sources; therefore, the decoherence process is described by two rates: the longitudinal relaxation rate $\Gamma_1 \equiv$ $1/T_1$ and the transverse relaxation rate $\Gamma_2 \equiv \frac{1}{T_2} \equiv \frac{\Gamma_1}{2} + \Gamma_\varphi$, where $T_1$ is the relaxation time, $T_2$ is the decoherence time and $\Gamma_\varphi$ is the pure dephasing rate. Typical superconducting qubits have a resonance frequency $\omega$ in the range between 3 and 6 GHz and are measured in a dilution refrigerator at a temperature $T \approx 20$ мК. Within these limits, due to the Boltzmann factor ($\exp(-\hbar\omega/k_B T)$) the qubit excitation rate is exponentially suppressed, and only relaxation contributes significantly to $T_1$.

Longitudinal relaxation occurs due to energy exchange between the qubit and its environment, causing a transition from the state $|1\rangle$ to $|0\rangle$. The primary sources of qubit relaxation are radiation into the control and readout lines due to the Purcell effect, as well as dielectric losses, which are largely driven by two-level systems, since the qubits operate in single photon regime. As with resonators, the energy dissipation rate due to TLS depends on the electric field strength in amorphous dielectrics, which is characterized by the participation ratio and depends on the gap between the qubit capacitor pads[168].

Pure dephasing is caused by noise that induces fluctuations in the qubit frequency $\omega$. This noise can exhibit a spectral dependence (such as charge noise, flux noise, or critical current noise). In some cases, charge noise driven by TLS coupled to the qubit becomes dominant[170]. There are also mechanisms of 'white' noise, typically arising from photon number fluctuations in the readout resonator coupled to the qubit[171].

To mitigate the noise sources and decoherence mechanisms described above, a specialized cryogenic environment is required[166]. Attenuation and filtering are essential for suppressing thermal noise propagating from the "hot" stages of the dilution refrigerator to the qubit. Attenuators distributed across various temperature stages attenuate the photon flux, ensuring that the effective noise temperature of the control lines matches the base temperature. Low-pass and infrared filters are used to block high-frequency radiation that could otherwise trigger quasiparticle generation, which significantly reduces the qubit's lifetime[167].

The microwave sample holder and shielding further protect the chip from external electromagnetic field. Multi-layer shields made of high-permeability materials (e.g., mu-metal) and superconductors prevent magnetic flux fluctuations. Additionally, the interior of the sample holder is often coated with microwave absorbers to suppress parasitic modes and stray radiation, which is critical for minimizing energy losses.

Superconducting qubit control is typically realized using microwave pulses delivered through microwave lines coupled to qubit. Manipulation of the qubit state occurs by applying resonant pulses at the qubit frequency $\omega$. By precisely tuning the amplitude, phase, and duration of these pulses, the qubit state vector can be rotated to any position on the Bloch sphere. For instance, a $\pi$-pulse is used to flip the qubit population from the $|0\rangle$ to state to the $|1\rangle$ state.

Qubit readout is most commonly performed via the dispersive measurement technique, where the qubit is coupled to a microwave readout resonator. The interaction causes a state-dependent shift in the resonator's frequency. By sending a probe signal to the resonator and analyzing the phase or amplitude of the reflected or transmitted wave, the qubit state can be determined non-destructively.

A detailed procedure for measuring qubit $T_1$ and $T_2$ is provided in review[172]. In a $T_1$ measurement experiment, the qubit is excited to the $|1\rangle$ state by a $\pi$ pulse, and its state is subsequently measured after varying time intervals. The probability of finding the qubit in the excited state decays over time. The final $T_1$ value is extracted from the exponential decay curve.

To determine $T_{2R}$ using the Ramsey method, two $\pi/2$ pulses are applied with a variable delay between them, driving the qubit into superposition and back. Due to external noise, the probability of the final state is described by decaying sinusoidal oscillations called Ramsey fringes. Data analysis involves fitting the envelope of this plot with an exponential function. The measurement of $T_{2E}$ via the Hahn echo method differs by the addition of a refocusing pulse exactly midway between the two $\pi/2$ pulses. The measured state probability is described by a pure exponential decay without oscillations. The resulting $T_{2E}$ time is often significantly longer than $T_{2R}$ because this method effectively filters out low-frequency noise, leaving only the impact of fast dynamical fluctuations.

### B. Performance of α-Ta resonators

We compiled published data on the fabrication and performance of α-Ta–based resonators, summarized in Table 2. The table includes substrate type and crystallographic orientation (where reported), substrate cleaning procedures, thin-film deposition temperature, and the presence and type of buffer layers. It further reports the resulting α-Ta crystallographic orientation, post-treatment methods, $p_{MS}$, $Qi_{LP}$, $Qi_{HP}$, and an estimated loss tangent tanδ ≈ δ. The latter is calculated as $\delta = (Qi_{LP} \times p_{MS})^{-1} \times 10^4$. We adopt the product $Qi_{LP} \times p_{MS}$ rather than $Qi_{LP} \times SPR$, as this quantity is directly reported in the literature[43,70,134]. For studies providing only geometric parameters (w/g) of the resonators, we model the participation ratios $p_{MS}$, $p_{MA}$ and $p_{SA}$, and subsequently compute the SPR. Across all analyzed devices, the SPR/$p_{MS}$ ratio lies in the range 1.9–2.7, indicating that SPR and $p_{MS}$ generally scale proportionally with the conductor width and gap. It should be noted that the calculated $p_{MS}$ values are approximate. In many cases, the etch depth into the Si substrate during RIE is not specified; however, the first ~100 nm of substrate ingress has the most significant influence on reducing $p_{MS}$. While SPR is commonly used to estimate dielectric losses via $\delta_{TLS}=1/Q_{TLS}=1/Qi_{LP}-1/Qi_{HP}$, this approach may not be reliable for Ta resonators. Specifically, $Qi_{HP}$ can be underestimated due to the premature onset of nonlinear behavior (discussed in the next chapter). Therefore, we estimate δ using $Qi_{LP}$. For clarity, the data are also visualized in Figure 4, which additionally indicates the α-Ta film growth method and RRR, where available.

*Table 2*

Performance of α-Ta resonators. Columns (from left to right): resonator number identifying the resonator in Figure 4 (№), reference where this measurement was reported (Ref.), deposition substrate (Substr.) indicating Si (100) or $Al_2O_3$ c-plane (c-pl.) or a-plane (a-pl.) orientation, substrate cleaning method (Clean), substrate temperature during deposition ($T_d$) and 2- to 15-nm-thick buffer layer used (Buf.), α-Ta crystallographic orientation (Orient. Ta), resonator chip post-treatment method (Post-treat.), calculated (thin font) and reported (bold font) metal-substrate interface participation ratio ($p_{MS}$), highest internal quality factor of the resonator at low power ($Qi_{LP}$) and at high power ($Qi_{HP}$), reported low-power loss $1/Qi_{LP}$ multiplied by $p_{MS}^{-1}$ (δ). Pir.: wafer treatment or chip post-treatment in Piranha solution. HMDS: wafer treatment with hexamethyldisilane. Ace.: wafer rinse in acetone. IPA: wafer rinse in isopropyl alcohol. BOE and HF: wafer treatment or chip post-treatment in hydrofluoric acid solutions. Degas: wafer heating to 300 °C for 12 hours. Descum: wafer surface cleaning with ions before film deposition. Ozone: ozone treatment. $AlO_X$ 2 nm: in-situ coating of α-Ta film with 2 nm $AlO_X$ layer. Ox.: Exposure of the resonators chip to air for several days after treatment in BOE. Long BOE: 120 min BOE post-treatment (2 to 20 min for all others).

| № | Ref | Substr. | Clean | Td,°C, Buf. | Orient. Ta | Post-treat. | $p_{MS}$ [$\times10^{-4}$] | $Qi_{LP}$ [$\times10^{6}$] | $Qi_{HP}$ [$\times10^{6}$] | δ [$\times10^{-4}$] |
|---|---|---|---|---|---|---|---|---|---|---|
| 1 | 67 | $Al_2O_3$, c-pl. | - | 500, - | 110 | - | | 0.20 | 1 | |
| 2 | -\|\|- | -\|\|- | - | 20, Nb | -\|\|- | - | | 0.07 | 0.2 | |
| 3 | -\|\|- | Si 100 | BOE | -\|\|- | -\|\|- | - | | | | |
| 4 | 69 | $Al_2O_3$, c-pl. | IPA, Degas | 530, - | 110 | - | **0.4** | 1.0 | 6 | 250 |
| 5 | -\|\|- | -\|\|- | -\|\|- | -\|\|- | -\|\|- | BOE | **0.4** | 5.0 | 60 | 50 |
| 6 | -\|\|- | -\|\|- | IPA, Descum | -\|\|- | -\|\|- | - | **0.4** | 2.0 | 35 | 125 |
| 7 | -\|\|- | -\|\|- | -\|\|- | -\|\|- | -\|\|- | BOE | **0.4** | 5.0 | 60 | 50 |
| 8 | 66 | Si 100 | BOE, HMDS | 600, - | 110 | BOE | | 2.10 | 8 | |
| 9 | -\|\|- | $Al_2O_3$, c-pl. | Pir. | -\|\|- | -\|\|- | Ozone | 9.2 | 3.0 | 6 | 3.6 |
| 10 | 151 | $Al_2O_3$ | | | 110 | Pir. | 8.4 | 3.0 | 10 | 4.0 |
| 11 | 144 | $Al_2O_3$, | | 500, Nb | | Pir. | 4.0 | 2.50 | 27 | 10.0 |
| 12 | 96 | $Al_2O_3$, c-pl. | Ace., IPA | 400, - | 110 | - | 13.0 | 0.67 | 27 | 11.5 |
| 13 | -\|\|- | -\|\|- | -\|\|- | 500, - | -\|\|- | - | 13.0 | 0.64 | 4.7 | 12.1 |
| 14 | 72 | $Al_2O_3$, c-pl. | Ace., IPA | 630, - | 110 | BOE | 15.4 | 0.45 | | 14.4 |
| 15 | -\|\|- | -\|\|- | -\|\|- | 630, Nb | 111 | -\|\|- | 13.8 | 2.0 | 20 | 3.6 |
| 16 | 70 | $Al_2O_3$, c-pl. | Pir., IPA | 500, - | 111 | - | **4** | 2.0 | 32 | 12.5 |
| 17 | -\|\|- | -\|\|- | -\|\|- | -\|\|- | 111 / 110 | - | **4** | 4.0 | 30 | 6.3 |
| 18 | 71 | $Al_2O_3$, c-pl. | | 625, - | 111 | Pir., BOE | 14.5 | 2.0 | 10 | 3.5 |
| 19 | 75 | Si 100 | HF 2% | 450, - | 110 | - | 7.9 | 1.10 | 54 | 11.4 |
| 20 | -\|\|- | -\|\|- | -\|\|- | -\|\|- | -\|\|- | HF 10% | 7.9 | 4.40 | 54 | 2.9 |
| 21 | 45 | Si 100 | | 500, - | | 500°C | 4.5 | 10.0 | 100 | 2.2 |
| 22 | -\|\|- | -\|\|- | | -\|\|- | | - | 4.5 | 2.50 | 100 | 8.9 |
| 23 | -\|\|- | -\|\|- | | -\|\|- | | HF 10% | 4.5 | 5.0 | 100 | 4.5 |
| 24 | 74 | Si 100 | BOE | 20, Nb | 110 | - | 7.5 | 0.60 | 15 | 22.1 |
| 25 | -\|\|- | -\|\|- | -\|\|- | -\|\|- | -\|\|- | HF 2% | 7.5 | 1.10 | 20 | 12.1 |
| 26 | 99 | Si 100 | BOE | 350, - | 110 | BOE | 2.2 | 6.25 | 20 | 7.2 |
| 27 | -\|\|- | -\|\|- | -\|\|- | -\|\|- | -\|\|- | -\|\|- | 14.9 | 3.80 | 33 | 1.8 |
| 28 | -\|\|- | -\|\|- | -\|\|- | 20, Nb | -\|\|- | -\|\|- | 2.2 | 9.09 | 33 | 5.0 |
| 29 | -\|\|- | -\|\|- | -\|\|- | -\|\|- | -\|\|- | -\|\|- | 14.9 | 2.85 | 25 | 2.4 |
| 30 | 137 | Si 100 | Ace., IPA | 20, Nb | | - | 33.7 | 0.45 | 0.8 | 6.6 |
| 31 | 109 | Si | | 20, TiN | | - | 5.6 | 2.10 | 3.8 | 8.4 |
| 32 | 135 | $Al_2O_3$, a-pl. | Pir. | 700, - | 110 | - | | 1.35 | 3.6 | |
| 33 | -\|\|- | -\|\|- | -\|\|- | -\|\|- | -\|\|- | $AlO_x$ 2 nm | | 1.08 | 3.7 | |
| 34 | 43 | Si 100 | Pir., HF 5% | 650, - | 110 | Pir., BOE | **2.0** | 10.0 | 200 | 5.0 |
| 35 | 107 | Si 100 | RCA, HF | 20, Nb | 110 | - | 12.4 | 1.0 | 10 | 8.0 |
| 36 | -\|\|- | -\|\|- | -\|\|- | -\|\|- | -\|\|- | BOE, Ox. | 12.4 | 1.20 | 5 | 6.7 |
| 37 | -\|\|- | -\|\|- | -\|\|- | 20, TaN | 110 | - | 12.4 | 0.40 | 2.5 | 20.1 |
| 38 | [44] | Si 100 | Pir., HF 2% | 500, - | 110 | Pir. | 3.7 | 3.0 | 8 | 9.0 |
| 39 | -\|\|- | -\|\|- | -\|\|- | -\|\|- | -\|\|- | Pir., BOE | 3.7 | 10.0 | 24 | 2.7 |
| 40 | [134] | $Al_2O_3$, | Pir., IPA | 750, - | 111 | Pir. | **7.8** | 0.80 | 2 | 16.0 |
| 41 | -\|\|- | -\|\|- | -\|\|- | -\|\|- | -\|\|- | Pir., BOE | **7.8** | 0.90 | 3 | 14.2 |
| 42 | -\|\|- | -\|\|- | -\|\|- | -\|\|- | -\|\|- | Pir., Triacid | **5.4** | 1.10 | 7 | 16.8 |
| 43 | -\|\|- | -\|\|- | -\|\|- | -\|\|- | -\|\|- | Pir., Long BOE | **4.9** | 4.0 | 100 | 5.1 |
| 44 | [76] | Si 100 | Ace., IPA, BOE | 600, - | 110 | BOE | 9.4 | 1.0 | 6.2 | 10.6 |
| 45 | -\|\|- | -\|\|- | -\|\|- | 20, TaN | 110 | -\|\|- | 9.4 | 0.14 | 0.5 | 75.7 |
| 46 | -\|\|- | -\|\|- | -\|\|- | 20, TiN | 110 | -\|\|- | 9.4 | 0.50 | 3.1 | 21.2 |
| 47 | [97] | $Al_2O_3$, a-pl. | - | 550, - | 110 | - | 8.4 | 1.25 | 2.4 | 9.5 |
| 48 | [141] | $Al_2O_3$, | | | | - | 6.2 | 0.75 | 10 | 21.6 |
| 49 | [68] | $Al_2O_3$, c-pl. | Pir. | -266, - | 110 | BOE | 14.6 | 1.50 | 5 | 4.6 |
| 50 | -\|\|- | Si 111 | Pir. | -\|\|- | -\|\|- | -\|\|- | 11.2 | 4.0 | 12 | 2.2 |

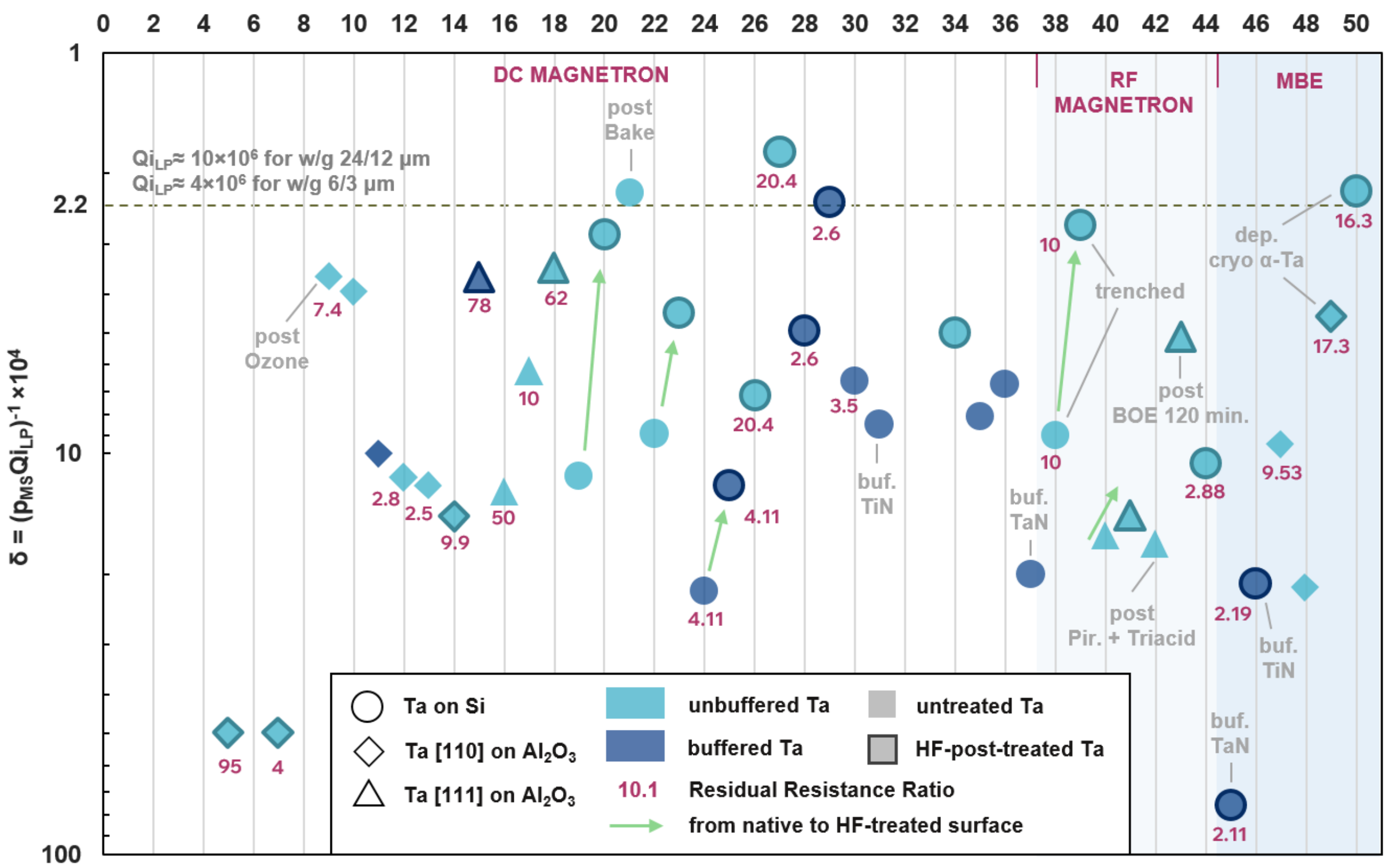

*Figure 4*

Reported low-power losses $1/Qi_{LP}$ multiplied by $p_{MS}^{-1}$, δ, for the resonators from Table 2 Circles refer to (110) α-Ta on Si. Diamonds refer to (110) α-Ta on $Al_2O_3$. Triangles refer to (111) α-Ta on $Al_2O_3$. The bold border of the symbol indicates Ta deposition on the buffer layer. The open symbol indicates a resonator that was treated in HF solution immediately before mounting in the cryostat for characterization. Red numbers report the residual resistance ratio reported for corresponding α-Ta films. Green arrows show the change in losses in identical resonators after treatment in HF solution. Blue legends at the top report the method of α-Ta film formation. Gray legends draw attention to the features of specific resonators (details in Table 2)

Figure 4 allows us to draw several conclusions. First, record-low losses were obtained for α-Ta films on Si treated in HF (for durations from 2 to 20 minutes), with the exception of post-annealed resonators (№ 21)[45]. Furthermore, a significant increase in the Q-factor of identical α-Ta on Si resonators can be seen after treatment in HF solutions (shown by green arrows in Figure 4). However, a 20-minute treatment in BOE of α-Ta resonators on $Al_2O_3$ provided almost no increase in the Q-factor (№ 40 and 41), since almost no tantalum oxide was removed during this time[134]. Therefore, the removal of native silicon oxide immediately before cryogenic characterization is crucial for increasing $Qi_{LP}$. Long-term (120 minutes) post-treatment of α-Ta resonators on $Al_2O_3$ in BOE also increased the $Qi_{LP}$[134]. However, there is evidence of a negative effect of long-term exposure to HF solutions on the Q-factor of α-Ta resonators due to the formation of non-superconducting tantalum hydride[45], which was also previously observed for niobium[173]. The aforementioned post-annealing of α-Ta resonators[45] appears promising for increasing the $Qi_{LP}$ of resonators, but further research and confirmation are required. A comparison of resonators № 35 and № 36 shows[107] that the recovery of native Si oxide after post-treatment in HF solution eliminates its positive effect. Secondly, the crystal orientation of Ta does not affect losses. Epitaxial α-Ta (111) on $Al_2O_3$ c-plane does not show an advantage over α-Ta (110) on Si and $Al_2O_3$. It should be noted that a direct correlation between RRR and losses is not observed, as was found for Nb[174], although a higher RRR indicates a smaller number of defects within the film. Third, the performance of heated α-Ta-based resonators is comparable to that of Ta/Nb-based resonators, but buffered Ta with TiN and TaN buffer layers exhibits increased losses. It is difficult to draw conclusions about the influence of the thin film formation method on the resonator losses, since there are few examples of using RF sputtering and MBE, but it should be noted that the very low losses have been shown for α-Ta grown at cryogenic temperature by the MBE method[68].

It should be emphasized that the loss values reported in Table 2 and Figure 4 are derived from the maximum reported $Qi_{LP}$ values (or, where necessary, extracted from graphical data). Within a single study, the variation in $Qi_{LP}$ for nominally identical resonators can reach several tens of percent. Moreover, the loss tangent δ used for comparison is inherently conditional and approximate. Accordingly, the present analysis does not aim to provide a rigorous quantitative comparison among different Ta-based resonators, but rather to establish an approximate, qualitative overview of the range of results reported in the literature.

The performance of α-Ta-based CPWRs is worth discussing separately compared to other key materials for superconducting circuits, such as Al and Nb. A direct comparison of the Qi of similar trench Ta and Al resonators was performed in work[44]. Although Ta-resonators post-treated in HF exhibited $Qi_{LP}$ greater than $10\times10^6$, a more relevant comparison would be between Ta resonators without post-treatment, with $Qi_{LP} \approx 3\times10^6$, and Al resonators, with $Qi_{LP} \approx 2\times10^6$, since Al is not compatible with post-treatment in HF solutions. Thus, the use of α-Ta provided a 1.5-fold reduction in low-power losses compared to Al. We are not aware of any comparison of α-Ta and Nb in a single study, but the best results separately for planar (50 nm etchings into the substrate)

Nb on Si CPW resonators with w/g 20/10 μm were shown in work[152]. For Nb CPWRs without post-treatment in BOE, $Qi_{LP} \approx 1.4\times10^6$, with a short 30 sec post-treatment removing only silicon oxides, $Qi_{LP} \approx 2.5\times10^6$, and with a long 20 min post-treatment in BOE removing also most of the Nb oxides, $Qi_{LP} \approx 6.0\times10^6$. For similar planar α-Ta-based CPWRs on Si with a w/g of 24/12 μm, $Qi_{LP} \approx 2.5\times10^6$ without post-treatment and $Qi_{LP} \approx 5.0\times10^6$ with a short post-treatment removing only Si oxides were shown in work[45]. However, the best $Qi_{LP}$ values for post-treated α-Ta resonators reach $10\times10^6$ (for slightly lower SPR) [43–45,99]. Thus, α-Ta resonators with removed native Si oxide provide at least a twofold advantage in low-power performance over similar Nb-based resonators, and replacing Nb with α-Ta without post-treatment provides an advantage of about 1.8 times.

### C. Performance of α-Ta qubits

The complexity of experiments[175] and the progress in error correction code efficiency[176,177] for superconducting qubits are determined by the number of qubits in the circuit and the error rates of quantum operations. The dominant portion of the error budget in quantum logic gates is attributed to the limited coherence of the qubits[176,177]. Thus, the performance improvement of quantum processors will depend on advancements in increasing qubit coherence times while simultaneously maintaining the scalability of the architecture and fabrication technology for systems with a larger number of qubits.

Due to the low losses at low microwave powers and cryogenic temperatures demonstrated in α-Ta resonators, as well as its chemical resistance to aggressive fabrication processes, tantalum has emerged as a promising material platform for qubit fabrication. The first implementations of qubits with an α-Ta base layer on sapphire substrates[41,42,178] achieved $T_1$ exceeding 300 us and quality factors $Q = 2\pi f_{qubit}\cdot T_1$ (which normalizes the relaxation time to the qubit frequency) up to $1\times10^7$. In subsequent studies, α-Ta was deposited on high-resistivity silicon substrates[179], which are expected to have lower bulk losses compared to sapphire[180] and are the preferred material for scalable manufacturing[181,182]. The transition to this new substrate enabled the demonstration of record lifetimes exceeding 1 ms and quality factors up to $1{,}5\times10^7$ for individual qubits[43]. In that work α-Ta was deposited via magnetron sputtering with argon onto a substrate heated to 450–650°C.

Further developments are aimed at creating a technology compatible with scalable semiconductor Back End of the Line (BEOL) processes, which require substrate heating not to exceed 400°C, to avoid damaging previously deposited aluminum structures[100]. Qubits based on α-Ta, deposited via magnetron sputtering on silicon using krypton at 350°C[100], as well as β-Ta[183], deposited at room temperature on silicon wafers, have demonstrated quality factors exceeding $1\times10^7$. A comparison with resonator performance indicates that qubit lifetimes are primarily limited by TLS-losses and could be further improved by removing surface dielectrics, for example, through encapsulation techniques[157].

However, it should be noted that the reported record relaxation times and quality factors were obtained on single qubits, which typically feature a large gap between the capacitor pads to minimize the contribution of amorphous interface dielectrics to total losses. This approach is not always compatible with scalable architectures that require a high density of interconnects. Furthermore, these 'hero devices' did not undergo subsequent fabrication steps, such as the formation of air-bridges[184] and flip-chip bonding[185], which can introduce contaminants and additional losses but are essential for manufacturing large-scale devices. Therefore, despite impressive results and a superiority in quality factors over other platforms under similar conditions[16,186–188], further refinement of technological processes is necessary, focusing on reducing dielectric losses in conjunction with full-scale integration. When evaluating results for scalable qubit technology, attention should be shifted from individual 'hero devices' toward the statistics across multiple qubits, a trend that has recently gained more focus from researchers. We have summarized the results for single Ta qubits in Table 3, presenting both the metrics for the best individual devices and the average values for larger sets of characterized qubits (where available). Data on the gap size, as well as coherence times from Hahn echo ($T_{2E}$) and Ramsey ($T_{2R}$) experiments, are also provided for comparison.

*Table 3*

Performance data for individual Ta-based qubits. All presented qubits utilize a transmon architecture. The 'Gap' column indicates the minimum distance between the qubit capacitor pads. Td denotes the substrate temperature during deposition. 'Best time-average' columns represent data averaged over measurement time for the single qubit device with the highest quality factor. 'Device-average' columns show data averaged across a set of multiple single-qubit devices.

| Ref | Substrate | Td, °C | Gap, um | Best time-average Q×10$^{-6}$ | Device-average Q×10$^{-6}$ | Best time-average T1, us | Device-average T1, us | Best time-average T2e, us | Device-average T2e, us | Best time-average T2R, us | Device-average T2R, us |
|---|---|---|---|---|---|---|---|---|---|---|---|
| 41 | $Al_2O_3$ | 500 | 70 | 7.1 | 5.5±1 | 303±32 | 240±51 | 201±28 | 100±61 | 105±3 | 65±24 |
| 42 | $Al_2O_3$ | -\|\|- | 80 | 11.7 | 8.0±2.6 | 476 | 302±104 | 463 | 336±97 | -\|\|- | 167±98 |
| 178 | $Al_2O_3$ | 800 | -\|\|- | 10.4 | -\|\|- | 280 | -\|\|- | 238 | -\|\|- | 62 | -\|\|- |
| 43 | Si | 600 | 70 | 15 | 9.7 | 1000±230 | 450±140 | -\|\|- | 450±370 | 170±56 | 103±37 |
| 100 | Si | 350 | 20 | 14.4 | 12.5±3 | 730±255 | 673±73 | 80±15 | 100±33 | 17±4 | 15±5 |
| 183 | Si | RT | 70 | 10.1 | 5.6 | 585±75 | 264±173 | 328±54 | 214±84 | 136±67 | 84±56 |

To date, results for multi-qubit Ta-based processors have been presented in only a few studies. A processor consisting of 63 transmon qubits and 105 couplers was fabricated using flip-chip technology[189]. It demonstrates moderate lifetimes, averaging 33 us, and was used for the experimental study of large-scale quasiparticle bursts induced by cosmic rays. Since the aluminum films of the Josephson junctions were in contact with the Ta films of the qubit capacitor plates, they acted as quasiparticle (QP) traps due to their smaller superconducting gap, thereby accumulating QPs. This led to an increased quasiparticle density in the aluminum regions and, consequently, to an elevated tunneling rate and a recombination time that was 2-3 orders of magnitude shorter than in a similar experiment on the Sycamore processor[190], which is fully made of aluminum.

In another study, a 13-qubit quantum processor[164], also fabricated using flip-chip technology, demonstrated a median relaxation time exceeding 100 us. Notably, the processor implements low-loss Ta airbridges that are resistant to aggressive chemical treatments used for contaminant removal (an option that is unavailable for Al-based airbridges).

In addition to the described technological challenges related to contaminants, air-bridges, and flip-chip assembly, other fundamental physical constraints become significant as the coherence of Ta-based qubits continues to improve. A recent study[191] has highlighted a specific characteristic of tantalum qubits: their increased sensitivity to background radiation. The results indicate that without additional infrared filtering, their coherence may be limited by a particle tunneling rate $\Gamma_0$ of up to 1.5 kHz. Consequently, as coherence times further increase, greater attention must be paid to filtering and shielding alongside the refinement of materials and fabrication processes.

### D. Advantages of α-Ta circuits

The use of α-Ta in superconducting resonators and qubits has been demonstrated to enable record-level performance. A key question concerns the physical mechanisms responsible for the reduction of low-power losses in α-Ta–based circuits. While post-treatment in HF solutions is known to be effective, the removal of native Si and Ta oxides – similar to the case of Nb-based circuits[192] – occurs only transiently, as rapid reoxidation follows. Consequently, for more complex circuit architectures incorporating not only an α-Ta base layer but also JJ and air bridges, such post-treatment becomes ineffective due to surface reoxidation[107,152]. Notably, as discussed in Section IV.B, α-Ta circuits fabricated on Si substrates exhibit superior performance compared to Al- and Nb-based counterparts even in the absence of post-treatment. This observation indicates that the improved performance of α-Ta–based devices is not solely attributable to oxide removal processes. Instead, it arises from a combined reduction of losses at both the substrate–metal (SM) and metal–air (MA) interfaces.

The source of dielectric loss at the SM interface is the transition layer between the substrate and the film, which arises due to the presence of contaminants and adsorbates on the substrate before film deposition and due to the interdiffusion of the substrate and film material. The state of the interfaces is typically assessed using transmission electron microscopy (TEM). For sapphire substrates, the SM interface is always sharp for both Ta/$Al_2O_3$[71,72,97,128] and Ta/Nb/$Al_2O_3$[71,72,127], and even when the $Al_2O_3$ substrate is heated to 1150°C, there is no intermixing with Ta[73]. At the same time, even at room temperature, Ta/Si[113] intermixing occurs during tantalum deposition on silicon, and during deposition on a heated substrate, the transition layer thickness is estimated to range from 4-5 nm[75,99] to 8-10 nm[44,71]. The Nb buffer layer in the Ta/Nb/Si structure also creates a transition layer with Si of approximately 3-4 nm thickness[74,99]. However, as can be seen from Figure 4, it is precisely on silicon substrates that record-breaking $Qi_{LP}$ values for α-Ta CPWRs were obtained. Thus, the contribution to the overall losses from the SM interface is not decisive.

The source of dielectric loss at the MA interface is the natural or, in some cases, technological oxide of the base layer material, i.e., a combination of some amount of amorphous main oxide ($Ta_2O_5$, $Nb_2O_5$, or $Al_2O_3$) and suboxides[193–197]. Therefore, the difference in the performance of Ta-, Nb-, and Al-based circuits can be associated with different structures and different TLS contents in their native oxides. The thickness of the native Ta oxide is not so small and, according to various estimates, ranges from 2 to 3 nm[73,151,198], 3 nm[43,72,74,97,128], 4 nm[45,133,136], 5 nm[99,158]. A comparison of native Ta and Nb oxides in one study[158] using electron energy loss spectroscopy (EELS) and X-ray photoelectron spectroscopy (XPS) studies showed that Ta oxide consists almost entirely of amorphous $Ta_2O_5$ (suboxide layer no more than 0.8 nm), while Nb oxide shows a gradual transition from $Nb_2O_5$ to $NbO_2$ and then to NbO as one moves deeper into the film (suboxide layer up to 1.6 nm). In addition, the study presents Ta-O and Nb-O phase diagrams obtained using the CALculation of PHAse Diagrams (CALPHAD) method, from which it follows that Ta has one stable stoichiometric oxide $Ta_2O_5$, while Nb has three oxides $Nb_2O_5$, $NbO_5$ and NbO, which is confirmed by other studies[133,152,159,199].

In work[200], using the highly sensitive method of surface chemical analysis variable energy X-ray photoelectron spectroscopy (VEXPS), $Ta_2O$ and $Ta_2O_3$ were detected in the native Ta oxide, but their thicknesses were only 0.37 and 0.37 nm, respectively. However, the main oxide $Nb_2O_5$ makes the main contribution to the low-power losses for Nb CPWRs[152,192,201,202], therefore, it can be assumed that $Nb_2O_5$ makes a greater contribution to the TLS losses than $Ta_2O_5$. This is to some extent confirmed by the results of work[203], where the losses in $Nb_2O_5$ and $Ta_2O_5$ deposited with different thicknesses on the Nb CPWR surface using pulsed laser deposition (PLD) were compared. The dielectric losses for $Nb_2O_5$ were 30% higher than for $Ta_2O_5$.

It has been shown[158] that amorphous $Ta_2O_5$ retains a more octahedrally symmetric crystalline character than $Nb_2O_5$. Hence, $Ta_2O_5$ has a reduced probability of oxygen ion tunneling between the two lowest-energy states, as predicted for disordered materials, which is equivalent to a smaller number of TLS defects compared to $Nb_2O_5$[174,204]. In other words, niobium oxide contains a higher number of oxygen vacancies acting as TLS[199,205]. DFT-calculations confirm[133,159] that the $Ta_2O_5$ surface contains fewer dangling oxygen bonds than the $Nb_2O_5$ surface. Moreover, DFT-simulations coupled with ab initio molecular dynamics (AIMD) showed[206] that at a given degree of oxygen deficiency, magnetic moment formation is suppressed in $Ta_2O_{5-x}$ compared to $Nb_2O_{5-x}$, eliminating hyperfine states that can act as resonant or subresonant TLS. Furthermore, $NbO_X$ may contain sources of non-TLS magnetic losses[207]. In addition to the lower TLS density in Ta oxide compared to Nb oxide, it has been shown[159] that the higher atomic mass of Ta compared to Nb shifts the TLS resonance frequencies in the native oxide below the gigahertz range in which qubits operate. The more ordered structure of $Ta_2O_5$ also makes it a promising candidate for the tunnel barrier in JJ[198]. Furthermore, the less distorted structure of $Ta_2O_5$ may more effectively block the diffusion of hydrogen atoms into the metal film through the native oxide, while hydrogen ion tunneling may also be a source of microwave losses[208,209], and hydrogen diffusion into the JJ barrier results to uncontrolled changes in its properties[210,211]. In Section II.C, we mentioned that the morphology of α-Ta on silicon without buffer layer includes smooth regions and regions with elongated, codirectional grains (ridges). According to terahertz scattering-type scanning near-field optical microscopy (s-SNOM), a boson peak is observed in the smooth regions[129], indicating a high content of amorphous $TaO_X$ in the surface oxide. This means that smooth regions potentially contain more TLS than ridged regions. Thus, the low low-power losses in Ta circuits are likely due to the low content of gigahertz-TLS in the native Ta oxide. Further improvement in the performance of Ta circuits requires long-term reduction of losses at the SA interface.

## V. Current issues in α-Ta-circuits

The effectiveness of tantalum as a base-layer material for superconducting circuits is well established. However, given its relatively recent adoption for this application, several aspects of its implementation remain insufficiently understood. Four key issues can be identified: the optimal method for Ta film formation, the influence of β-Ta impurities, the effect of HF solutions on the properties of Ta films, and the origin of the nonlinear behavior observed in α-Ta CPWRs with increasing signal power or temperature.

In Sections II.A and III.B, we demonstrated that there are four approaches to forming α-Ta thin films and at least five deposition methods. As shown in Figure 4, CPWRs fabricated from films deposited by magnetron sputtering on heated substrates and those grown by MBE on cryogenic-temperature substrates exhibit comparable performance. Furthermore, both films deposited without a buffer layer and those grown on buffer layers achieve record-level characteristics. These observations indicate that, at present, it is not possible to unambiguously identify an optimal approach or deposition method for α-Ta film formation, and this issue warrants further systematic investigation.

### A. Influence of β-phase impurities

At first glance, it may be expected that films consisting purely of the α phase would exhibit optimal properties, since the β phase is generally considered unfavorable for superconducting applications. However, it was shown[69] that the presence of 10% β-phase in the form of isolated regions at the MS interface has a beneficial effect on the performance of resonators and also leads to an increase in the critical magnetic field for the film. Furthermore, it was found that an α-Ta film on Si can be grown with the preliminary formation of a thin (≈10 nm) β-Ta underlayer[44,130]. Moreover, resonators fabricated from such films demonstrate record-breaking performance[44]. In work[76], heated α-Ta resonators also showed a higher $Qi_{LP}$ than buffered α-Ta resonators, despite the presence of β-Ta in the film, according to XRD data. In the case of α-Ta growth on a $TiN_x$ buffer layer, increasing the substrate temperature resulted in a higher β-phase fraction, while simultaneously increasing the RRR of the film[110]. These observations indicate that the presence of β-Ta does not necessarily degrade device performance and may, under certain conditions, be advantageous. Therefore, it cannot currently be concluded that β-Ta should be entirely eliminated from α-Ta films, and this issue requires further investigation.

A recent study[183] presented transmon qubits and CPWRs fabricated from β-Ta with $T_C \approx 0.7$ K on $Al_2O_3$ substrate, with the qubit Q-factor reaching $10\times10^6$, corresponding to a $T_1$ of 585 μs. The surface TLS losses for β-Ta are estimated to be twice as high as for similar α-Ta-based devices, based on the characterization of CPWRs with various geometries, and the films also exhibit significant kinetic inductance. However, this work suggests that the presence of minority β-Ta in α-Ta films may not be as detrimental to the performance of superconducting circuits as previously assumed, and hence the stringent requirement for "phase-pure α-Ta" deposition can be relaxed. The potential for using β-Ta in superconducting qubits as well as kinetic inductance detectors requires further study.

### B. Effect of hydrofluoric acid solutions on α-Ta

Hydrofluoric acid solutions remove native silicon oxide, temporarily passivating the surface, and partially etch native tantalum oxide. As shown in Section IV.B, HF post-treatment plays a key role in achieving record-level performance of α-Ta CPWRs. However, it has also been demonstrated[45] that prolonged exposure to HF can have a detrimental effect on CPWR performance. Specifically, exposure to 10% HF for 1 - 2 minutes results in an increase in $Qi_{LP}$, whereas longer exposure leads to a severe degradation of both $Qi_{LP}$ and $Qi_{HP}$. This behavior has been attributed to the formation of tantalum hydride in the film, which suppresses superconductivity; notably, this effect can be mitigated by annealing at 500 °C for 1 hour (Figure 5). In the same study, the etch rates of $Ta_2O_5$ and Ta in 10% HF were reported to be 0.4 nm/min and 15 nm/min, respectively. In contrast, other studies[75] employing 10% HF limited post-treatment durations to no more than 1 minute, while CPWR circuits treated in 10:1 BOE solutions were exposed for 20[43,107] and even 120 minutes[134], and in 7:1 BOE solutions for up to 15 minutes[69,76], without apparent degradation in performance. These observations suggest that each HF-based solution is characterized by a critical exposure time, beyond which active formation of tantalum hydride occurs, leading to a sharp decline in CPWR performance. Additionally, for heated α-Ta films on Si, we observe a pronounced suppression of $Qi_{LP}$ and $Qi_{HP}$ following post-treatment for 1 minute in a 50% HF solution. Thus, the determination of an optimal post-treatment duration for α-Ta circuits in HF-based solutions remains an open question, requiring a balance between maximal removal of oxide layers and avoidance of tantalum hydride formation.

### C. Anomalous behavior of α-Ta resonators

The most intriguing effect in the behavior of α-Ta CPWRs, in our opinion, is the nonlinearity at high-power signal (greater than $10^3$–$10^6$ photons) and the radical decrease in Qi at measurement temperatures above 100 mK. CPWR nonlinearity at high powers consists of a sharp decrease in Qi when a certain signal power is exceeded (Figure 5). In some studies, this decrease is clearly shown in the plots[45,69,134], while in other studies, the dependence of the CPWR Q-factor on the signal power simply terminates prematurely[44,71,72,74,97]. This effect occurs for both sapphire[69,97] and silicon[44,45] substrates. Note that we are not aware of any nonlinearity for buffered α-Ta CPWRs, with the exception of Ta/TiN resonators with TSV-terminated membranes[109]. At the same time, a sharp decrease in Qi is detected as CPWRs heat up to 100 mK[69,134], while other α-Ta CPWRs maintain performance up to 700 mK[69,145,148] (Figure 5). The CPWR resonant frequency also decreases abnormally rapidly with increasing temperature and signal power[45,69] (Figure 5), and the measured data for the $S_{21}$ scattering parameter in the IQ-plane exhibit an elliptical shape[45] (Figure 5). What is particularly striking is that CPWR circuits fabricated on the same wafer can exhibit different behavior[69,71,134] (Figure 5)!

In work[45], the nonlinearity of α-Ta CPWRs was explained by a two-photon loss mechanism[212] related to the heating of quasiparticles due to hydrogen impurities[213] after post-treatment in an HF solution. Indeed, since the scattering parameter $S_{21}$ in the IQ plane does not retain a circular shape, dissipative nonlinearity can be assumed rather than the effect of kinetic inductance[214]. In work[69], nonlinearity also occurs after post-treatment of the CPWR circuit in an HF solution; however, in other works[44,70,97] nonlinear behavior was reported for CPWRs that were not exposed to HF.

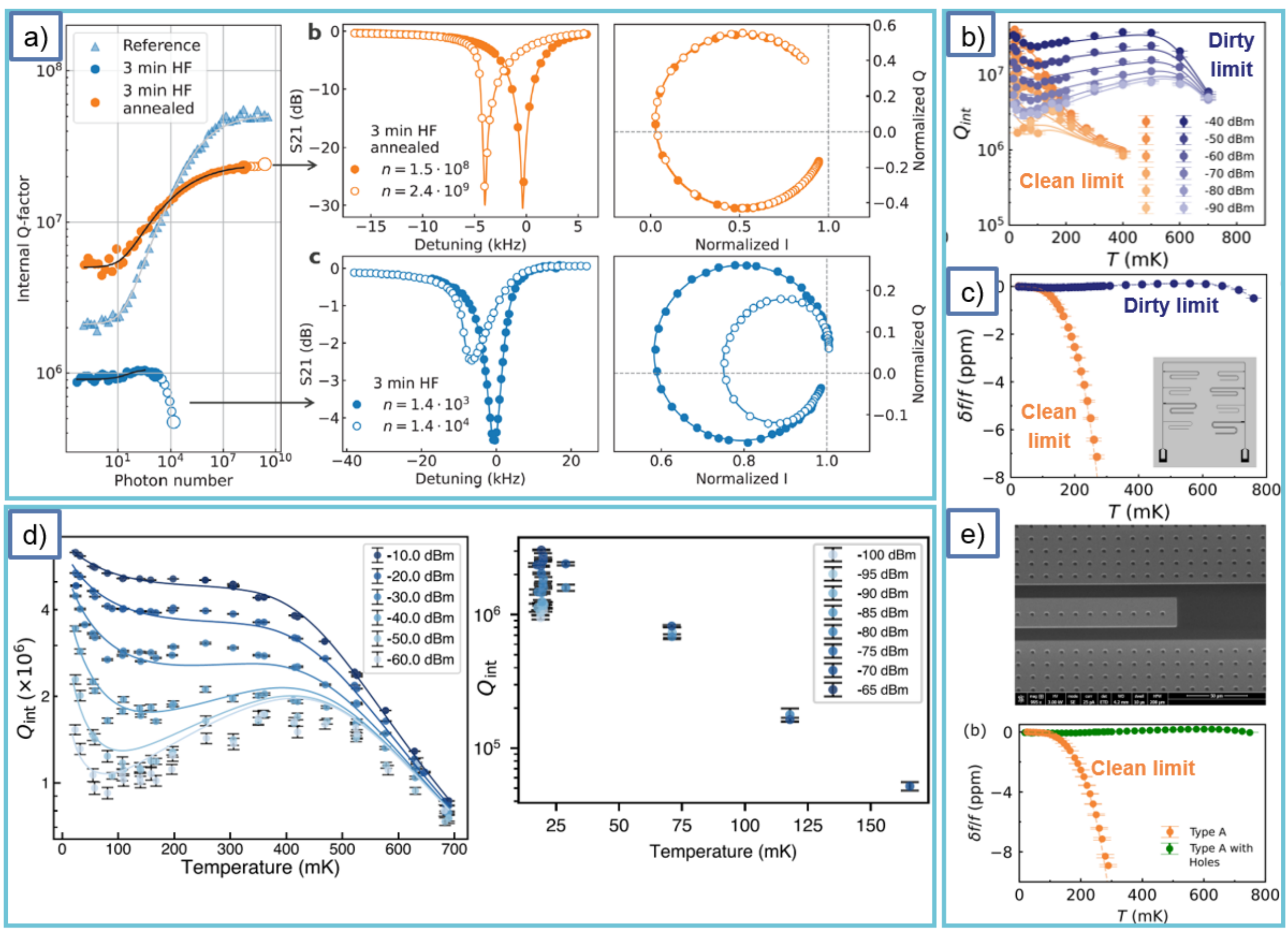

*Figure 5*

Nonlinear behavior of α-Ta resonators. a) Photon-number-dependent internal Q-factor for resonators subjected to different treatments; scattering parameters as a function of frequency for a sample dipped in HF and subsequently annealed; scattering parameters as a function of frequency for a sample dipped in HF without annealing treatment (the elliptical shape in the IQ plane is visible), reprinted from[45], under a Creative Commons license (https://creativecommons.org/licenses/by/4.0/). b) Measured internal Q-factor as a function of temperature with varying applied microwave power for resonators made from films in the clean (orange) and dirty (blue) limit (a sharp decrease in Q-factor is visible for the clean limit). c) Measured relative frequency shift as a function of temperature for resonators made from films in the clean (orange) and dirty (blue) limit (a sharp decrease in frequency is visible for the clean limit), adapted from[70], Copyright (2025) American Physical Society. d) Internal Q-factor of resonators fabricated from a single wafer as a function of applied microwave power and temperature; the left plot corresponds to the typical resonator behavior with a sharp decrease in Qi for temperatures above 400 mK, and the right plot corresponds to the nonlinear behavior with a sharp decrease in Qi at minimal heating, reprinted from[134], under a Creative Commons license (https://creativecommons.org/licenses/by/4.0/). e) SEM image of a resonator fabricated from a film in the pure limit with trap holes introduced to suppress vortex motion and the frequency shift as a function of temperature for a resonator fabricated from a film in the pure limit without holes (orange) and with holes (green), adapted from[70], Copyright (2025) American Physical Society.

The most comprehensive study to date of the nonlinear behavior of α-Ta CPWRs is presented in work[70]. It was found that α-Ta films, which are used to fabricate CPWRs exhibiting nonlinear behavior, are in the pure superconducting limit, where the coherence length of superconductivity is shorter than the mean free path. Such films have few sites for pinning magnetic vortices, so vortex motion results to energy dissipation. In films in the dirty limit, defects ensure vortex pinning and impede their motion. The film type was determined based on the magnetic field strength required to suppress superconductivity. For films in the dirty limit, the strength was twice as high. In addition, for such films, the RRR did not exceed 12, while for films in the clean limit, the RRR ranged from 38 to 62, which also confirms a higher number of structural defects for films in the dirty limit. Local areas of the β-phase could be assumed as a site for pinning in the case of films in the dirty limit, but they were not detected even by scanning SQUID microscopy. XRD data show that all α-Ta films in the clean limit have a (111) orientation when grown on $Al_2O_3$ c-plane, which means heteroepitaxial growth, while films in the dirty limit can have both orientations (110) or (111). The authors associate the film type with its deposition regime and propose patterning the CPWR centerline and grounding to create artificial vortex pinning sites, similar to flux trapping holes, as a method for protecting against vortex motion[215] (Figure 5). The authors' evidence in ref.[70] appears convincing; however, in their earlier work[134], they reported that CPWRs with normal and nonlinear behavior can be present within a single wafer (i.e., the film deposition regime is the same). Similarly, refs.[44,69] report different behavior of resonators obtained from a single crystal. Furthermore, ref.[44] shows that α-Ta films exhibit RRR≈10 and (110) orientation, i.e., they are in the dirty limit; however, nonlinearity is present. Therefore, at a minimum, the question remains of reliably identifying the film type before a CPWR circuit or qubit circuit is fabricated, since creating flux trapping holes directly on a coplanar line will inevitably lead to increased dielectric losses in the circuit[215]. Thus, fully unraveling the causes of the nonlinear behavior of α-Ta-based circuits is one of the most important challenges in α-Ta superconducting circuit technology.

## VI. Conclusions and Outlook

Thus, α-Ta currently represents the most promising material for the base layer of superconducting circuits, including qubits and coplanar resonators. This conclusion is supported by the recent demonstration of millisecond-coherence transmon qubits based on tantalum. The principal advantages of α-Ta over Al and Nb include its high chemical resistance, which enables aggressive post-treatment of the circuit base layer, and the composition and structure of its native oxide, characterized by a reduced density of TLS within the relevant frequency range. At the same time, several fundamental and technological questions remain unresolved. In particular, the mechanism governing phase selection in tantalum thin films has not yet been conclusively established, and therefore the optimal approach to α-Ta film formation and the most suitable deposition method remain unclear. Although record-level performance of coplanar resonators has been achieved using magnetron sputtering with substrate heating, devices fabricated by molecular beam epitaxy on cryogenic-temperature substrates exhibit comparable characteristics. An additional advantage of α-Ta is its excellent performance on silicon substrates, which are standard in the microelectronics industry. In this context, the chemical resistance of tantalum is especially important, as it permits processing in hydrofluoric acid solutions that remove native $SiO_2$ and temporarily passivate the silicon surface for a period of time sufficient to load the circuit into a cryostat for measurements. However, there is also evidence that hydrofluoric acid can negatively affect circuit performance under certain conditions, and this factor must be carefully considered. Another open question concerns the role of β-Ta inclusions. While α-phase purity might intuitively be expected to yield superior performance, several studies report a beneficial effect of β-Ta regions, as well as high performance in resonators incorporating thin β-Ta underlayers. Additionally, structural defects in α-Ta films may play a positive role by promoting the pinning of magnetic vortices. Finally, a major outstanding challenge in the field of α-Ta–based circuits is the origin of the anomalous nonlinear behavior observed with increasing signal power or operating temperature, which appears intermittently across samples. Although several plausible mechanisms have been proposed, this phenomenon remains insufficiently understood and warrants further investigation.

Judging by the growing number of publications devoted to α-Ta–based superconducting circuits, interest in this material is steadily increasing, and further significant developments addressing existing open questions can be expected in the near future. It is therefore reasonable to anticipate that tantalum will emerge as a primary material platform for superconducting qubits and resonators, potentially displacing aluminum and niobium in many applications. In addition, early efforts toward the development of $TaO_x$-based Josephson junctions with a tantalum oxide barrier indicate the possibility of extending this material system to all key elements of superconducting circuits. Such progress may ultimately enable a transition toward fully tantalum-based device architectures, representing a new stage in the evolution of superconducting circuit technology.

## AUTHOR DECLARATIONS

### Conflict of Interest

The authors have no conflicts to disclose.

### Funding Declaration

This research received no specific grant from any funding agency in the public, commercial, or not-for-profit sectors.

### Author Contributions

**Evgeniy V. Zikiy**: Conceptualization (lead); Data curation (lead); Formal analysis (lead); Investigation (lead); Validation (lead); Visualization (equal); Writing – original draft (lead). **Nikita S. Smirnov**: Data curation (equal); Formal analysis (equal); Investigation (equal); Validation (equal); Writing – original draft (equal). **Stanislav A. Kotenkov**: Data curation (equal); Formal analysis (supporting); Validation (supporting). **Ilya A. Stepanov**: Visualization (equal); Formal analysis (supporting). **Julia A. Agafonova**: Data curation (equal); Validation (supporting). **Daria A. Moskaleva**: Formal analysis (equal); Writing – review & editing (supporting). **Aleksei R. Matanin**: Data curation (equal); Writing – review & editing (supporting). **Anton I. Ivanov**: Investigation (supporting); Writing – review & editing (supporting). **Dmitry O. Moskalev**: Investigation (supporting); Writing – review & editing (supporting). **Julia A. Kurochkina**: Investigation (supporting); Writing – review & editing (supporting). **Aleksander V. Andriyash**: Project administration (lead); Supervision (lead); Writing – review & editing (equal). **Ilya A. Rodionov**: Conceptualization (equal); Project administration (lead); Supervision (lead); Writing – review & editing (lead).

## DATA AVAILABILITY

The data that support the findings of this study are available from the corresponding author upon reasonable request.